\documentclass[]{interact}

\usepackage{epstopdf}
\usepackage[caption=false]{subfig}

\usepackage[numbers,sort&compress,merge]{natbib}
\bibpunct[, ]{[}{]}{,}{n}{,}{,}

\theoremstyle{plain}

\theoremstyle{definition}

\theoremstyle{remark}

\usepackage{mathtools}

\begin{document}


\title{Laser Cooling of Trapped Ions in Strongly Inhomogeneous Magnetic Fields}

\author{
\name{Richard Karl, Yanning Yin and Stefan Willitsch\thanks{CONTACT S. Willitsch. Email: stefan.willitsch@unibas.ch}}
\affil{Department of Chemistry, University of Basel, Klingelbergstrasse 80, 4056 Basel, Switzerland}
}

\maketitle

\begin{abstract}
Hybrid traps for the simultaneous confinement of neutrals and ions have recently emerged as versatile tools for studying interactions between these species at very low temperatures. Such traps rely on the combination of different types of external fields for the confinement of either species raising the question of interactions between the individual traps. Here, the influence of a strongly inhomogeneous magnetic field used for trapping neutrals on the trapping and laser cooling of a single Ca$^+$ ion in a radiofrequency ion trap is studied theoretically using molecular-dynamics simulations based on multilevel rate equations. The inhomogeneous magnetic field couples the different components of the ion motion and introduces position-dependent Zeeman splittings. Nonetheless, laser cooling is still found to work efficiently as the ion samples different magnetic field strengths and directions along its trajectory. Offsetting the centres of the two traps generates a linear magnetic-field gradient so that multiple lasers are required to address the resulting range of Zeeman splittings in order to ensure efficient cooling. The present study yields detailed insights into the ion cooling dynamics in combined magnetic and radiofrequency electric fields relevant for the characterisation and optimisation of hybrid trapping experiments.
\end{abstract}


\section{\label{sec:Intro}Introduction}
Magnetic ($B$-) fields are widely used tools for the manipulation of the external as well as the internal degrees of freedom of ions. For instance, in Penning traps homogeneous $B$-fields with strengths 1-10 T introduce Lorentz forces that confine charged particles perpendicularly to the $B$-field while static electric fields trap them along the $B$-field direction \cite{thompson09a}. At the same time, the $B$-field defines the quantisation axis of the internal energy levels of the ions by lifting degeneracies via the Zeeman effect. In contrast to the strong magnetic fields in Penning traps, radio-frequency (RF) traps confine ions only using electric fields, and the quantisation axis is typically defined by an additional weak homogeneous $B$-field of strength 1-10 G. The difference in the field strengths requires different laser-cooling strategies for the two types of traps. The Zeeman splittings caused by fields of a few Gauss are often smaller than the laser linewidths used in Doppler cooling such that a single laser beam can, in principle, drive all Zeeman sublevels of an optical cycling transition of an ion in a RF trap. In Penning traps, the Zeeman splittings are significantly larger than the laser linewidths and more elaborate schemes involving additional lasers are necessary \cite{koo04a} to address all relevant states. Moreover, RF traps are typically designed such that the ion motion is harmonic and all motional modes can be efficiently cooled by a single laser beam at an angle to the three principal axes of the trap \cite{neuhauser78a}. The strong magnetic field in a Penning trap also causes a cyclotron and a magnetron motion. Additional procedures like offsetting the laser beams from the trap centre \cite{itano82a} or applying an oscillating field to couple the cyclotron and the magnetron motion \cite{powell02a} are required to cool all motional modes in a Penning trap.\\
Moreover, magnetic fields are sometimes intentionally added to engineer the level structure of particles. For instance, when working with molecules or exotic matter an adequately chosen homogeneous $B$-field can mix dark states and thereby improve the laser cooling efficiency \cite{kozyryev17a,lim18a,zimmer21a}.\\
Advanced laser cooling schemes involving $B$-field gradients have been studied theoretically for improved cooling efficiency in the Lamb-Dicke regime \cite{albrecht2011a} and for cooling of molecular anions via a magnetically-assisted Sisyphus process \cite{yzombard15a}. $B$-field gradients are also widely used in trapped-ion quantum-science experiments. Both static \cite{wang09b,johanning09a,khromova12a,weidt16a} and oscillating \cite{ospelkaus11a,warring13a,harty16a,hahn19a,zarantonello19a,srinivas21a} field gradients have been used to realise laser-free entangling gates and individual ion addressing. In those experiments, relatively small gradients of about 0.25-25~T/m were used and no significant effect on the laser cooling efficiency was reported. \\
A higher degree of inhomogeneity is present in quadrupolar magnetic fields which form the basis of magnetic and magneto-optical traps. There are numerous theoretical and experimental studies of neutral atoms and molecules in quadrupolar magnetic traps \cite{migdall85a,weinstein98a,perez-rios13a} and in magneto-optical traps \cite{raab87a,shuman10a,tarbutt15a,devlin16a}. Such traps, however, are usually not used for ions since ion traps already provide deep and stable trapping. Nevertheless, understanding the motion of ions in quadrupolar magnetic fields and how their light-matter interaction is affected by the $B$-field becomes relevant with the development of hybrid traps. In hybrid traps, neutral species and ions are trapped simultaneously which allows the study of cold and ultracold ion-neutral collisions with long interaction times \cite{willitsch15a,tomza19a}. Hybrid traps have been realised by overlapping RF ion traps with magnetic \cite{zipkes10a}, optical \cite{schmid10a,meir16a,joger17a} or magneto-optical traps \cite{smith05a,grier09a,hall11a,rellergert11a} for neutrals. \\
The success of these experiments with neutral atoms inspires similar studies with trapped neutral molecules to be combined with ions. However, magnetic traps for molecules usually exhibit much stronger inhomogeneous fields than those used for atoms \cite{haas19a}, raising the question of the influence of the strong $B$-fields on the ion motion and laser-cooling dynamics. In this manuscript, we report molecular-dynamics simulations to explore laser cooling of an ion in a strong quadrupolar magnetic field. The insights gained in the present study provide a more detailed understanding of hybrid trapping experiments and may serve towards their optimisation and extension to molecular species.

\begin{figure}[!h]
    \centering
    \includegraphics[width=\linewidth]{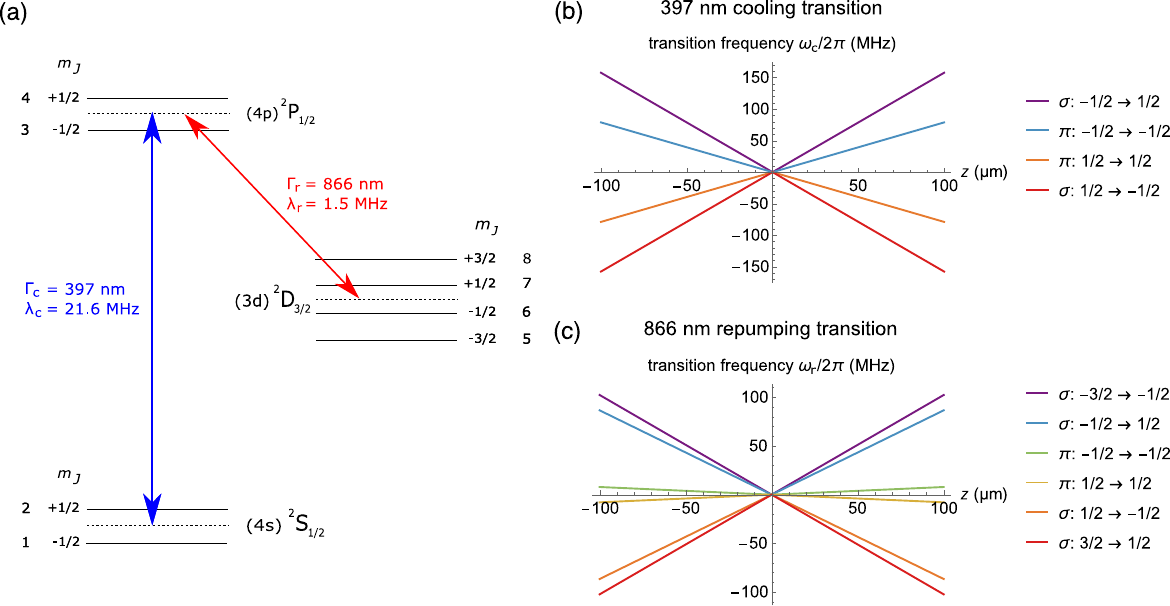}    
    \caption{(a) Energy level structure relevant for the Doppler laser cooling of $^{40}$Ca$^+$. The field-free fine-structure levels are indicated by dotted lines with transition wavelengths $\lambda_t$ and decay rate constants $\Gamma_{t}$ ($t=$c/r for cooling/repumper) according to reference \cite{safronova11a}. In a $B$-field, these levels split into an eight-level system. Throughout this manuscript, the Zeeman sublevels are numbered from 1 to 8 as indicated in the figure. (b,c) Frequencies of $\sigma$- and $\pi$-transitions between Zeeman components (indicated by their magnetic quantum numbers $m_J$) of the cooling (b) and repumper (c) transitions as a function of the axial ($z$) position of the ion in the hybrid trap. }
    \label{fig:Calcium}
\end{figure}

\section{\label{sec:Model}Theoretical models}
\subsection{\label{sec:DC} Model for Doppler laser cooling}
In Doppler laser cooling, photon scattering from a red-detuned laser beam generates a directional force that cools the motion of a counter-propagating atom \cite{foot05a}. The scattering rate, and thus the cooling force, depends on the internal properties of the atom via the spontaneous decay rates on selected spectroscopic transitions and on the external dynamics via the Doppler and Zeeman shifts and position-dependent laser parameters. In a hybrid trap composed of a RF trap and a magnetic trap, micromotion and magnetic field gradients can lead to timescales of the external dynamics that are comparable to the timescales of the internal population dynamics. Therefore, it is not valid to assume steady-state level populations of the ion as it is often done in laser cooling models based on a continuous friction force \cite{zhang07a,bell09a,okada10a}. \\
Instead, a stochastic model is more appropriately used for the light-matter interaction in which discrete transitions and discrete time steps are assumed \cite{rouse15a,marciante10a}. Such a model assumes that the ion is in one specific quantum state at a time which can be thus tracked over the course of the simulation. At each time step, the probability for an absorption or an emission event is computed. By comparing the absorption and emission probabilities to a random number between zero and one, it is decided whether an absorption or emission event takes place within a given time step $\delta t$. In such a case, the state of the ion is updated and the ion experiences a recoil either parallel or anti-parallel to the laser propagation direction, or in a random direction depending on whether absorption, stimulated emission or spontaneous emission occurred. For a particular state, the probability for emission is given by the sum of the spontaneous-decay and the stimulated-emission probabilities. The spontaneous decay probability is given by $\Gamma \delta t$ with $\Gamma$ being the natural linewidth of the relevant transition. In the present stochastic model, the stimulated decay probability is assumed to be equal to the probability for absorption which we express as $R \delta t$ (see also section \ref{sec:RE}). Here, $R$ is the excitation rate. By choosing $\delta t \ll 1/R$ and $\delta t \ll 1/\Gamma$ it is ensured that the absorption and emission probabilities are smaller than 1 at each time step \cite{rouse15a}. \\
Besides Doppler cooling, a stochastic model naturally captures processes that are either neglected or require further elaboration in continuous force models. These include heating due to photon recoil, fluctuations of the absorption rate, stimulated emission and saturation effects \cite{rouse15a}.

\subsection{\label{sec:RE}Rate equations for state-population dynamics}
To implement the stochastic model for laser cooling, an expression for the excitation rate $R$ is needed. We use multilevel rate equations \cite{tarbutt15a,comparat14a,xu19a} to describe the internal dynamics of a trapped ion from which we derive an expression for $R$. The validity of the current rate-equation treatment was verified against optical Bloch equations as discussed in section \ref{sec:App_REvsOBE} in the appendix. In experiments using alkaline-earth ions such as Ca$^+$ \cite{willitsch12a}, Doppler laser cooling is implemented using a $^2$S$_{1/2} - ^2$P$_{1/2}$ cooling transition and a $^2$D$_{3/2} - ^2$P$_{1/2}$ repumper transition, see figure \ref{fig:Calcium}. \\
We distinguish between the cooling and the repumper transition via the index $t=\{\text{c},\text{r}\}$ and denote the decay rate constants as $\Gamma_{t}$. In a magnetic field, this three-level energy structure splits into an eight-level system (figure \ref{fig:Calcium}). Here, the index $j$ indicates Zeeman sublevels of the (4p)~$^2$P$_{1/2}$ state, while the Zeeman sublevels of the (4s)~$^2$S$_{1/2}$ and of the (3d)~$^2$D$_{3/2}$ states are referred to as $i$. When $i$ is a Zeeman sublevel of (4s)~$^2$S$_{1/2}$, $t$ is set to $\text{c}$. Conversely, $t=\text{r}$ when $i$ is a Zeeman sublevel of (3d)~$^2$D$_{3/2}$. Within this notation, the dynamics of the populations $N_j$ of the $^2$P$_{1/2}$ Zeeman sublevels is given by
\begin{equation}
\label{eq:N_j}
\Dot{N_j} = -\left(\Gamma_{\text{c}}+\Gamma_{\text{r}}\right) N_j +
\sum_{i,l_t} R_{i,j,l_t} (N_i-N_j),
\end{equation}
in which $R_{i,j,l_t}$ is the excitation rate constant for the transition $i\leftrightarrow j$ driven by the laser beam $l_t$. The evolution of $N_i$ is described by
\begin{equation}
\label{eq:N_i}
\Dot{N_i} = \sum_{j,l_t} R_{i,j,l_t} (N_j-N_i) + \sum_j r_{i,j} \Gamma_t N_j,
\end{equation}
in which $r_{i,j}$ is the branching ratio for the spontaneous decay of level $j$ to level $i$. The excitation rate constants can be expressed as a sum over laser beams $l_t$ that independently drive the transition $t$ \cite{tarbutt15a}:
\begin{equation}
\label{eq:Excitation_rate}
R_{i,j,l_t} = \sum_{l_t} \frac{f_{i,j,l_t} \ \Omega_{l_t}^2\ \Tilde{\Gamma}_{l_t}}{4\ \left( \delta_{l_t} - \Vec{k}_{l_t}\cdot\Vec{v} - \omega^\text{z}_{i,j} \right)^2 + \Tilde{\Gamma}_{l_t}^2}.
\end{equation}
Here, $\delta_{l_t} = \omega_{l_t} - \omega_t$ is the detuning of a laser beam with angular frequency $\omega_{l_t}$ from the transition angular frequency $\omega_t$, $\Vec{k}_{l_t}$ is the wave vector of laser beam $l_t$, $\Vec{v}$ is the velocity of the ion, and $\omega_{i,j}^\text{z} = (g_j m_j - g_i m_i) \mu_{\text{B}} B/\hbar$ is the Zeeman shift. $\mu_{\text{B}}$ is the Bohr magneton, $\hbar$ is the reduced Planck constant, $B$ is the magnetic field strength, $g_j$, $g_i$ are the relevant Land\'e $g$-factors and $m_j$, $m_i$ are the magnetic quantum numbers of the sublevels $j$ and $i$. We define $\Tilde{\Gamma}_{l_t} = \left(\Gamma_{\text{c}}+\Gamma_{\text{r}}\right) + L_{l_t}$ as the sum of the decay rate constants and the full-width-at-half-maximum (FWHM) linewidth $L_{l_t}$ of laser beam $l_t$. $\Omega_{l_t} = \Gamma_t \sqrt{S_{l_t}/2}$ is the Rabi frequency with which a laser with saturation parameter $S_{l_t}=I_{l_t}/I^{\text{sat}}_t$ drives the transition $t$ in absence of a $B$-field. The fractional strength $f_{i,j,l_t}$ of the transition $i\leftrightarrow j$ driven by laser beam $l_t$ with polarisation $\Vec{\epsilon}_{l_t}$ is given by 
\begin{equation}
\label{eq:frac_transition_str}
f_{i,j,l_t} = \frac{|\langle i | \Vec{D} \cdot \Vec{\epsilon}_{l_t} | j \rangle|^2}{\sum_{i} |\langle i|\Vec{D}|j \rangle|^2},
\end{equation}
where $\Vec{D}$ is the dipole moment operator. The quantisation axis of the ion is defined by the local $B$-field ("atomic frame") in which atomic properties like the dipole moment operator $\Vec{D}$ are typically expressed \cite{oberst99a}. The polarisation vector $\Vec{\epsilon}_{l_t}$ is defined relative to the $k$-vector of the laser beam $l_t$ and therefore needs to be transformed into the atomic frame to evaluate equation (\ref{eq:frac_transition_str}). This transformation needs to be computed at every time step of the ion trajectory (see section \ref{sec:EoM}) as the $B$-field is inhomogeneous.
\begin{figure}[!h]
    \centering
    \includegraphics[width=\linewidth]{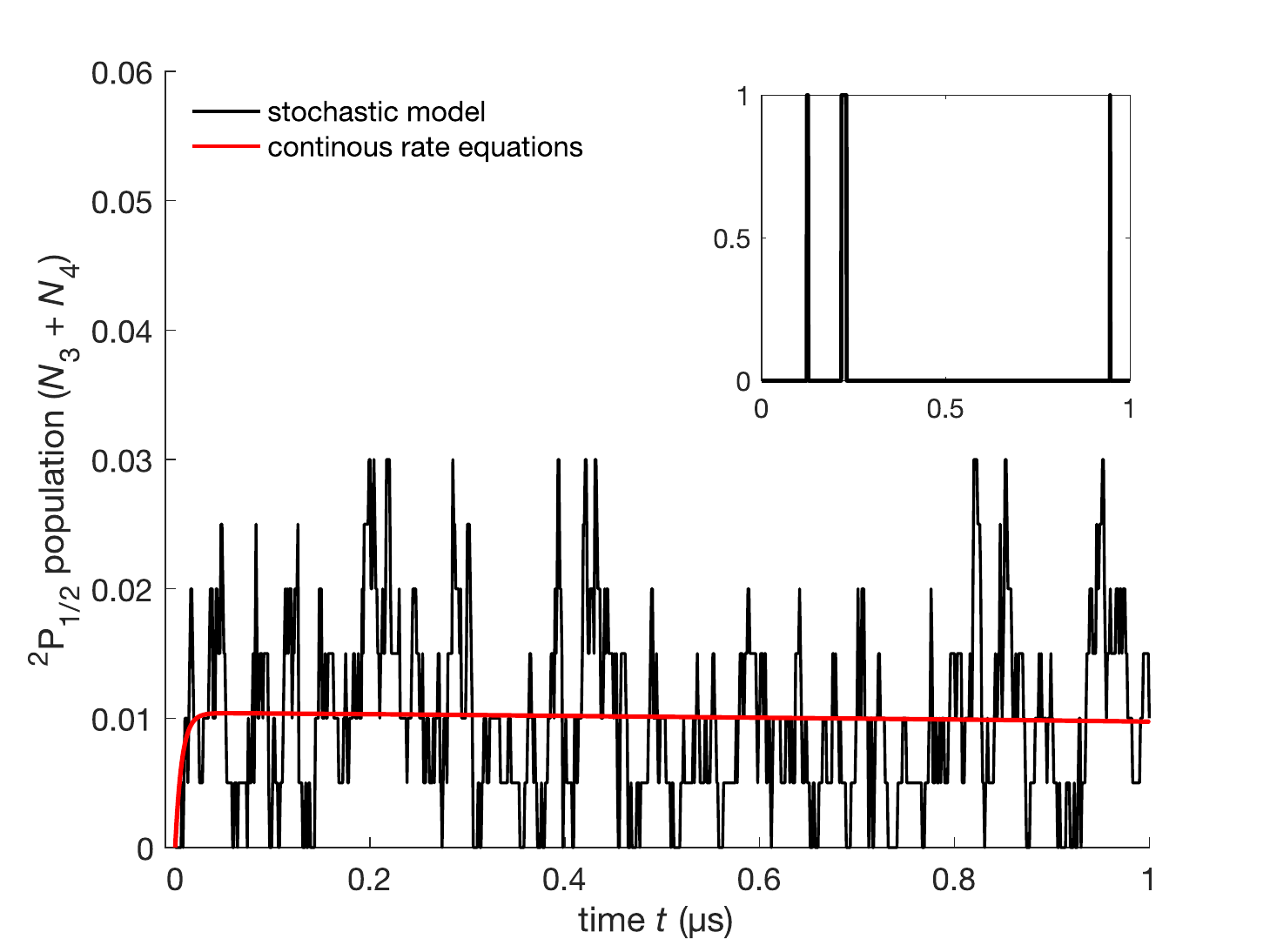}
    \caption{The sum of the populations of the two $^2$P$_{1/2}$ Zeeman sublevels $N_3$ and $N_4$ according to our stochastic model (black) based on eight-level rate-equations. The inset shows the result of the stochastic model for a single simulation run. The ensemble average over multiple repetitions (here 200 times in the black trace) approaches the continuous solutions of the rate equations (red trace). Two counter-propagating cooling and two counter-propagating repumper lasers with saturation parameters $S_{\text{c},1}=S_{\text{c},2}=S_{\text{r},1}=S_{\text{r},2}=10$, detunings $\delta_{\text{c},1}=\delta_{\text{c},2}=-30$~MHz, $\delta_{\text{r},1}=\delta_{\text{r},2}=-10$~MHz and left circular polarisation with respect to the laser propagation direction at a $B$-field of 100~G were assumed.}
    \label{fig:SateTracking}
\end{figure}\\
Since in our stochastic model an ion can only be in one state at a time (see section \ref{sec:Cooling}), $N_i$ and $N_j$ are either 0 or 1. Depending on the state of the ion, the excitation rate $R$ is therefore either zero or given by the excitation rate constant $R_{i,j,l_t}$ expressed in equation (\ref{eq:Excitation_rate}). According to the stochastic nature of the model, a single quantum jump only occurs very rarely and seemingly randomly throughout short simulation times. This is depicted in the inset of figure \ref{fig:SateTracking} where the population evolution of the $^2$P$_{1/2}$ state for a single calcium ion at rest is shown. When the same simulation is repeated multiple times, the population averaged over the ensemble approaches the continuous solution of the rate equations (\ref{eq:N_j}) and (\ref{eq:N_i}).

\subsection{\label{sec:EoM}Equations of motion for a trapped ion}
We assume that the motion of a single trapped ion in an inhomogeneous magnetic field and under laser irradiation is subject to three forces: the electric Lorentz force, the magnetic Lorentz force, and the scattering force. While the scattering force is treated via the stochastic model described above, the forces due to an electric field $\Vec{E}$ and a magnetic field $\Vec{B}$ acting an ion with mass $m$, charge $e$ and velocity $\Vec{v}$ are incorporated into Newtonian equations of motion,
\begin{equation}
\label{eq:Newton}
\frac{d\Vec{v}}{dt} = \frac{e}{m} \left( \Vec{E} + \Vec{v} \times \Vec{B}\right).\\
\end{equation} 
We assume that the ion is confined in a harmonic RF trap such that the electric field can be expressed as \cite{major05a},          
\begin{equation}
\Vec{E} = \frac{m \Omega_{\text{RF}}^2}{4 e} 
\begin{pmatrix}
(-a+2 q \cos(\Omega_{\text{RF}} \ t)) x \\ 
(-a-2 q \cos(\Omega_{\text{RF}} \ t)) y \\
2 a z \\
\end{pmatrix}.
\end{equation} 
Here, $x,y,z$ are the position coordinates of the ion and $a = -\frac{4 \kappa e V_{\text{dc}}}{m z_0^2 \Omega_{\text{RF}}^2}$ and $q = -\frac{2 e V_{\text{rf}}}{m r_0^2 \Omega_{\text{RF}}^2}$ are the Mathieu parameters which depend on the ion's charge $e$ and mass $m$, the geometric parameters $r_0$, $\kappa$, $z_0$ assuming a linear RF quadrupole trap \cite{willitsch12a}, the applied voltages $V_{\text{dc}}$ and $V_{\text{rf}}$ and the RF angular frequency $\Omega_{\text{RF}}$. An integration scheme that is adequate to numerically solve equation (\ref{eq:Newton}) is the Boris algorithm \cite{qin13a,birdsall18a,penn03a} which is a finite difference method with second-order accuracy \cite{penn03a}. It only requires one evaluation of the field per time step $\delta t$ and is stable for oscillatory motion as it conserves phase space volume \cite{qin13a}. However, in contrast to the widely used leapfrog scheme, the acceleration is allowed to be velocity-dependent which is necessary to model the magnetic Lorentz force \cite{qin13a}. Defining $t_k\coloneqq k\delta t$, $\Vec{x}_k\coloneqq \Vec{x}(t_k)$, $\Vec{v}_k\coloneqq \Vec{v}(t_k-\delta t/2)$, $\Vec{E}_k\coloneqq \Vec{E}(\Vec{x}_k)$ and $\Vec{B}_k\coloneqq \Vec{B}(\Vec{x}_k)$, the Boris algorithm can be implemented as \cite{birdsall18a}
\begin{equation}
\begin{split}
&\Vec{v}^{\,-} = \Vec{v}_{k} + \frac{q}{m} \Vec{E}_k \frac{\delta t}{2},\\
&\Vec{v}^{\,'} = \Vec{v}_{\,-} + \Vec{v}^{\,-} \times \Vec{t},\\
&\Vec{v}^{\,+} = \Vec{v}_{\,-} + \Vec{v}^{\,'} \times \Vec{s},\\
&\Vec{v}_{k+1}= \Vec{v}^{\,+} + \frac{q}{m} \Vec{E}_k \frac{\delta t}{2},\\
&\Vec{x}_{k+1}= \Vec{x}_{k} + \Vec{v}_{k+1} \delta t.\\
\end{split}
\end{equation}
The temporary variables $\Vec{v}^{\,-}$, $\Vec{v}^{\,'}$ and $\Vec{v}^{\,+}$ are defined in the first, the second and the third line of the algorithm and the quantities $\Vec{t}$ and $\Vec{s}$ are defined as
\begin{equation}
\Vec{t} = \frac{q \delta t}{2 m} \Vec{B}_k, \;\; \text{and} \;\;\; \Vec{s} = \frac{2}{1 + |\Vec{t}\,|^2} \Vec{t}.
\end{equation}

\subsection{\label{sec:Par}Simulation parameters}
Doppler laser cooling of a single calcium ion was studied in a hypothetical hybrid trap composed of a linear-quadrupole RF trap and a quadrupolar magnetic trap. For the simulations reported here, we used $\Omega_{\text{RF}} = 8$ MHz, $a = -3.18 \cdot 10^{-4}$ and $q = -0.08$ which correspond to realistic parameters of traps used in our laboratory \cite{vonPlanta20a}. The inhomogeneous $B$-field was assumed to originate from two bar magnets in anti-Helmholtz configuration with dimensions $3\times4\times6$ mm as used, e.g., in ref. \cite{haas19a}. Each magnet was assumed to have a remanence of 1.64 T and the surface-to-surface distance between the two bar magnets was taken as 3.6 mm. A sketch of the assumed setup and a simulation of the $B$-field are shown in section \ref{sec:App_ExpSetup} in the appendix.\\
Figures in this article show the average of 32 simulations with different initial velocity vectors of the ion and with different sets of random numbers used in the Monte Carlo model of the cooling force (see section \ref{sec:DC}). The ion was initialised in the (4s)~$^2S_{1/2}, m_j=-1/2$ state and the initial position of the ion was chosen at the centre of the hybrid trap. Initial velocity vectors of the ions were sampled over random directions with their norm satisfying the equipartition theorem such that the initial kinetic energy of the ion was 1~K. Such a scenario can experimentally be realised by Doppler-precooling the ion in the RF trap before it is overlapped with the magnetic trap. Choosing an initial ion position slightly displaced from the hybrid trap centre or assuming higher initial temperatures yielded qualitatively similar results (see section \ref{sec:App_MorePlots} in the appendix).\\
If not otherwise mentioned, we assume two Gaussian laser beams, one for cooling (subscript $\text{c}$) and one for repumping (subscript $\text{r}$), which propagate along the trap axis ($z$ axis in figure \ref{fig:ExpSetup} (b) in the appendix) with a power of 0.5 mW (corresponding to saturation parameters $S_{\text{c}} = 1.96$ and $S_{\text{r}} = 105.75$, respectively) and with $1/e^2$ waist radii of $W_{\text{c}} = 0.6$ mm and $W_{\text{r}} = 1$ mm. If not otherwise mentioned, we further assume laser detunings of $\delta_\text{c}=-30$ MHz and $\delta_\text{r}=-10$ MHz, FWHM linewidths of $L_{\text{c}} = L_{\text{r}} = 0.3$ MHz and linear polarisations which are parallel to the $x$ axis of the lab frame (see section \ref{sec:App_ExpSetup} in the appendix). Since a single ion is studied and only laser beams along the axial direction of the trap are assumed, cooling in only one dimension is obtained as discussed in the following sections. As in experiments without strong magnetic fields, three-dimensional cooling can be achieved by setting the direction of the cooling-laser beam at a finite angle to all three principal axes of the trap or working with larger ensembles of ions in which the Coulomb interaction couples all motional degrees of freedom \cite{willitsch12a}.

\section{\label{sec:Results}Results and Discussion}
\subsection{\label{sec:Lorentz}Ion motion in combined electric radio-frequency and magnetic fields}
A cold ion initialised close to the centre of the presently considered hybrid trap has velocity components on the order of tens m/s and samples $B$-field strengths less than 100 G (see section \ref{sec:App_ExpSetup} in the appendix). For a calcium ion, this results in a Lorentz force on the order of $10^{-19}$ to $10^{-20}$~N which is comparable to the magnitude of the scattering force. However, with laser cooling the Lorentz force becomes less pronounced over time as the velocity of the ion as well as its motional amplitude and thereby the magnitude of the experienced $B$-field decrease due to the cooling.\\
In the absence of a $B$-field, the motion of an ion in a linear RF trap is determined by the Mathieu equations \cite{major05a}. The resulting ion motion is a combination of a slow component, termed secular motion, and a fast oscillatory motion called micromotion. Micromotion originates from the RF fields and occurs therefore only along the radial directions of the trap, along which the RF fields are applied, but not the axial one. If a homogeneous $B$-field is applied, the Lorentz force causes the ion to undergo cyclotron and magnetron motions in the plane orthogonal to the $B$-field direction, especially at large field strengths. In a quadrupolar magnetic field, the $B$-field has a different magnitude and direction at every position. Cyclotron and magnetron frequencies therefore also vary with the position. Therefore, we do not expect to observe Fourier components with frequencies corresponding to well-defined cyclotron and magnetron motions in the present case. Instead, we expect the Lorentz force to cause a mixing of the radial and axial frequency components. Moreover, since the magnetic Lorentz force is small compared to the confining electric forces of the linear Paul trap, the secular and micromotions are expected to dominate. 
\begin{figure}[!h]
    \centering
    \includegraphics[width=\linewidth]{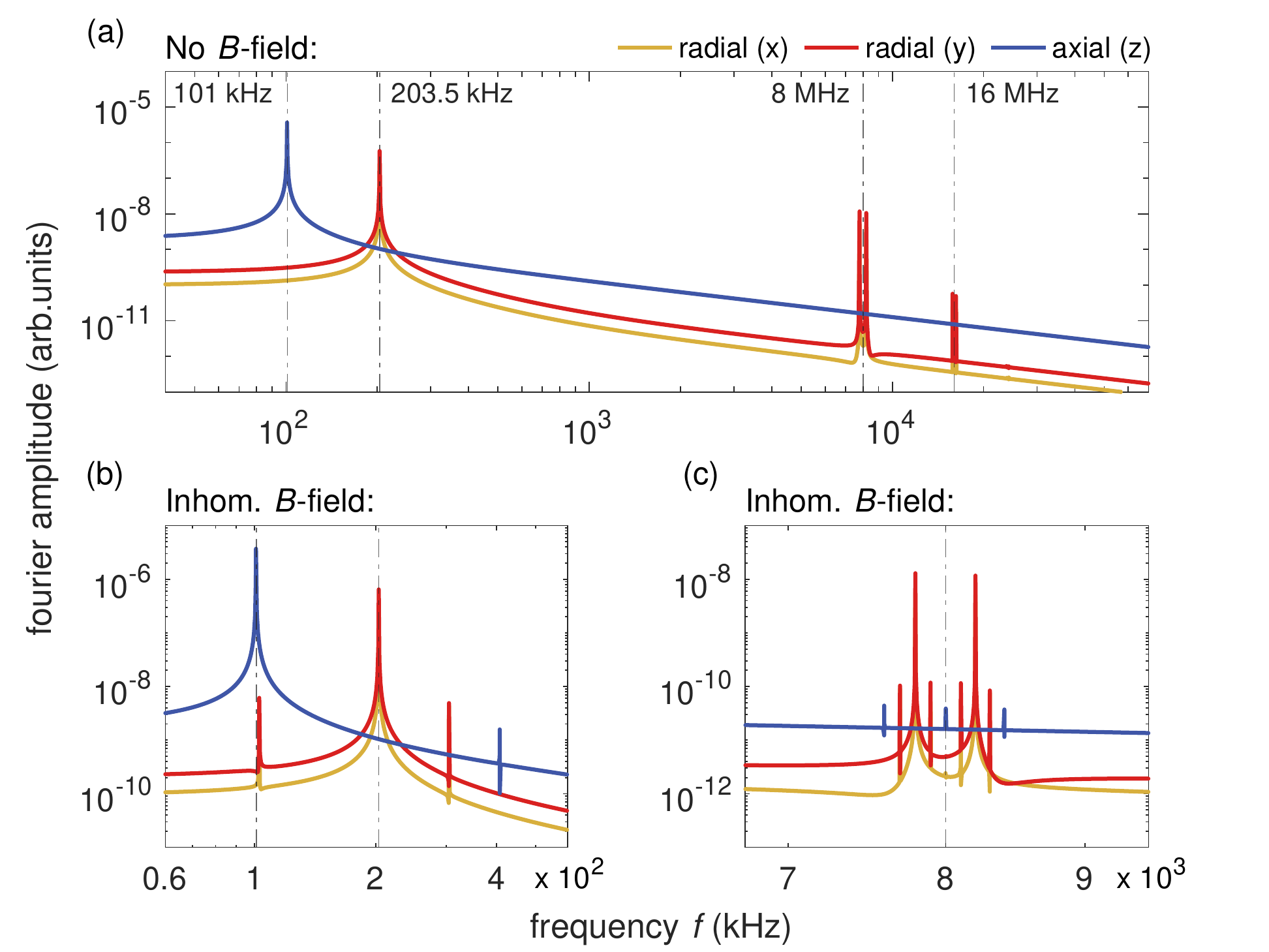}
    \caption{Fourier spectra of the motion of an ion trapped in a linear Paul trap (a) in the absence of a $B$-field and (b) \& (c) with an overlapped quadrupolar magnetic field. The dominant features correspond to the frequencies of the secular motion (101 kHz, 203.5 kHz), the micromotion (8 MHz), and its second harmonic (16 MHz). In the inhomogeneous $B$-field, the motions along different axes mix with each other and produce peaks in the Fourier spectrum at sum and difference frequencies.}
    \label{fig:Fourier_comp}
\end{figure}\\
Solving equation (\ref{eq:Newton}) numerically in the absence of laser beams confirms these expectations. Without a magnetic field, four dominant features are observed in the Fourier spectrum of the ion motion in figure \ref{fig:Fourier_comp}. The peaks at 101 kHz and 203.5 kHz correspond to the axial and radial secular motions, respectively. Since we assume an idealised ion trap, the secular frequencies of motion along the two radial axes are degenerate. The peaks around 8 MHz and 16 MHz correspond to the micromotion frequency and its second harmonic. The micromotion is modulated by the secular motion and thus consists of two peaks split by twice the radial secular frequency. In the hybrid trap, i.e., including the inhomogeneous $B$-field, the same secular and RF frequencies are present but additional Fourier components at 102.5 kHz, 304.5 kHz and 407 kHz appear which are the sum and difference frequencies between the secular frequencies along different axes (figure \ref{fig:Fourier_comp} (b)). Under closer inspection, additional peaks around 8 MHz can also be observed which correspond to the sum and difference frequency of the modulation of the micromotion by the secular motion (figure \ref{fig:Fourier_comp} (c)).

\subsection{\label{sec:Cooling}Laser-cooling dynamics of trapped ions in inhomogeneous $B$-fields}
An ion at 1~K oscillates around the trap centre with amplitudes of up to 40 $\umu$m and, 
\begin{figure}[!h]
    \centering
    \includegraphics[width=\linewidth]{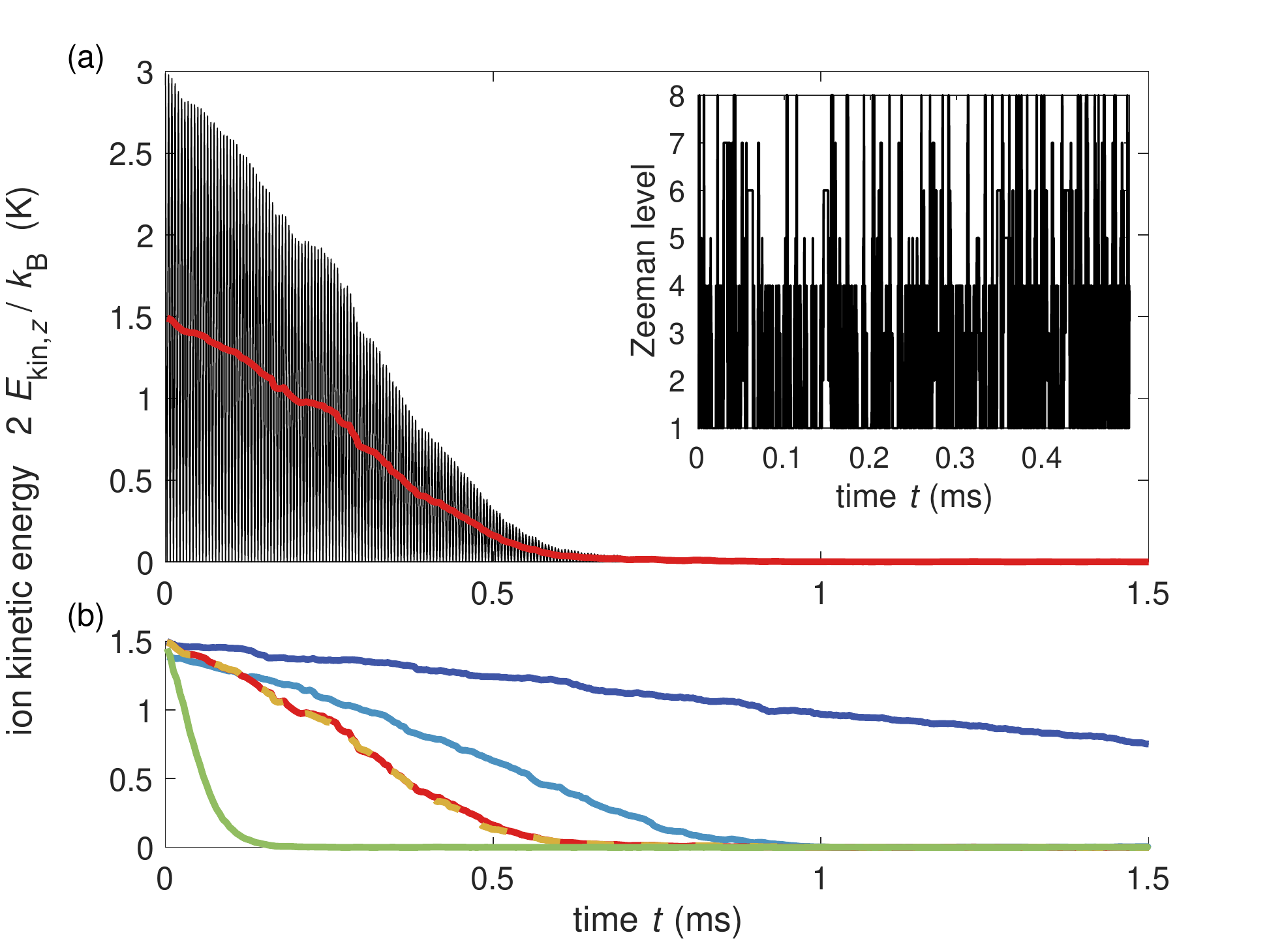}
    \caption{(a) Example of the time evolution of the kinetic energy of an ion along the axial direction (black trace) and its average over one period of the secular motion (red trace). The inset shows the population of the eight Zeeman sublevels over the course of the ion trajectory for the first 0.5 ms (see figure \ref{fig:Calcium} for labelling of the levels). (b) Averaged ion kinetic energies for different experimental configurations. Green trace: without magnetic field; red trace: cooling and the repumper lasers with linear polarisation and aligned with the $x$ axis (same situation as in panel (a)); yellow-dashed trace: linear polarisation aligned in the $\hat{x}+\hat{y}$ direction; blue trace: left circular polarisation; light-blue trace: left circular polarisation for an initial velocity vector with higher radial components ([4.80, -4.21, 24.11] m/s compared to [0.68, -0.89, -24.92]~m/s used in the other simulations). Laser detunings: $\delta_\text{c}=-30$~MHz, $\delta_\text{r}=-10$~MHz. }
    \label{fig:Tcurve_1c1r_P0000}
\end{figure}
thereby, samples $B$-field strengths that cause Zeeman splittings of up to 120~MHz (see figure \ref{fig:Calcium}). Despite the relatively large Zeeman splittings, efficient laser cooling can be achieved with one cooling and one repumper laser, as illustrated in figure \ref{fig:Tcurve_1c1r_P0000}. As a figure of merit for the cooling efficiency, we adopt the time $t_{1/2}$ required to remove half of the ion's initial kinetic energy along the axial direction (henceforth dubbed "half-life"). In each simulation, the kinetic energy (black trace in figure \ref{fig:Tcurve_1c1r_P0000} (a)) was averaged over one period of the secular motion and half-lives were then obtained from these averages (red trace in figure \ref{fig:Tcurve_1c1r_P0000} (a)). In figure \ref{fig:Dscan_1c1r_P0000}, the half-life is plotted for different laser detunings. 
\begin{figure}[!h]
    \centering
    \includegraphics[width=\linewidth]{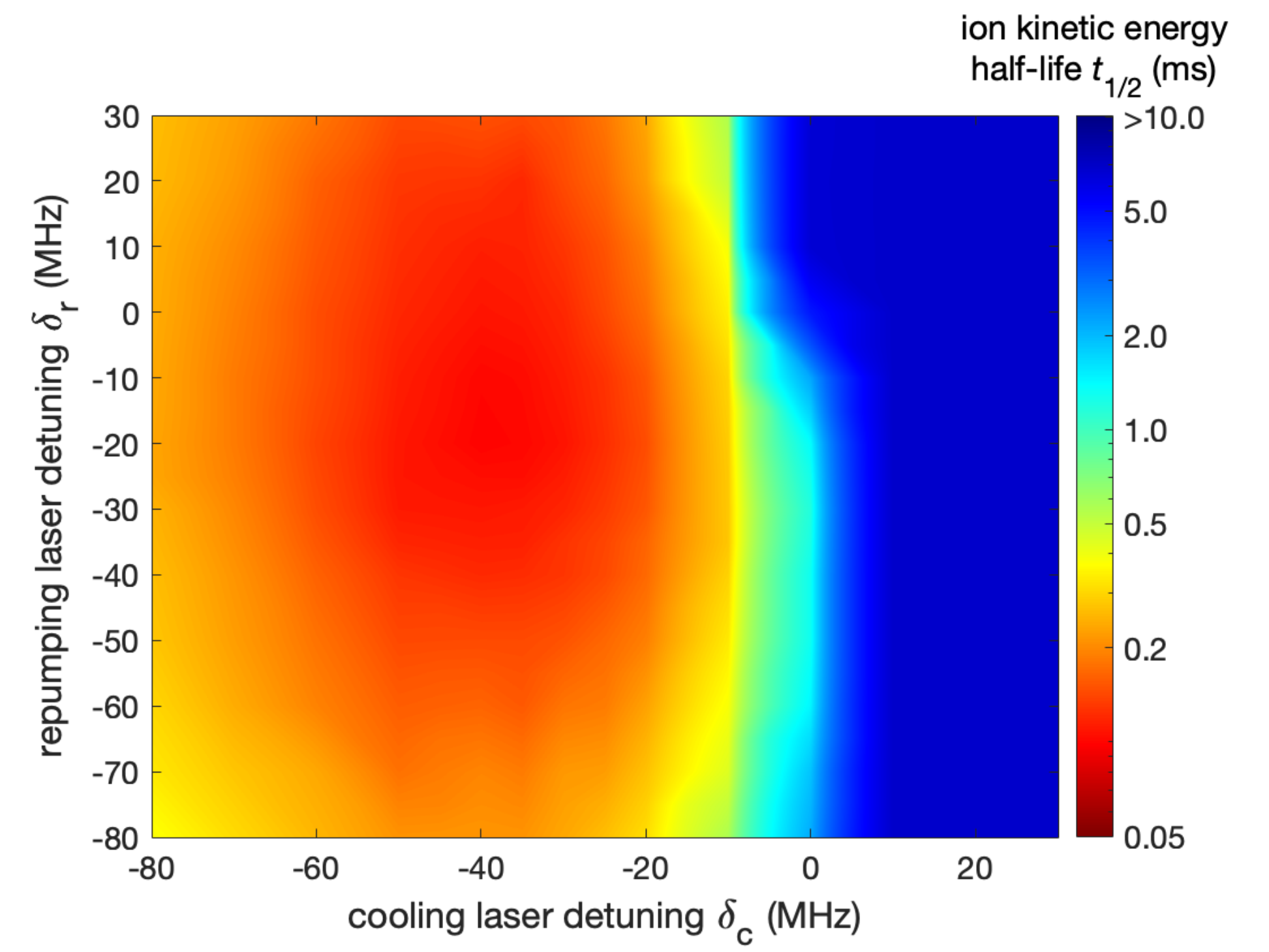}
    \caption{Ion kinetic energy half-lives $t_{1/2}$ (time required to remove half of the ion's kinetic energy) at different cooling-laser ($\delta_\text{c}$) and repumper detunings ($\delta_\text{r}$) for a single ion in the hybrid trap. Doppler cooling requires a red-detuned cooling laser beam ($\delta_\text{c}<0$). The shortest cooling half-lives are found between $\delta_\text{c}=-20$~MHz and $\delta_\text{c}=-60$~MHz and for repumper laser detunings in a range of more than a hundred MHz around the repumper transition resonance.}
    \label{fig:Dscan_1c1r_P0000}
\end{figure}\\
The fastest cooling is achieved when the cooling laser is detuned between -20~MHz and -60~MHz from the $^2$S$_{1/2} - ^2$P$_{1/2}$ transition frequency at zero $B$-field. The exact detuning of the repumper laser is less critical than the detuning of the cooling laser since most absorption events occur between the $^2$S$_{1/2}$ and the $^2$P$_{1/2}$ states. As required for Doppler cooling, the cooling laser must be red-detuned ($\delta_\text{c}<0$), otherwise heating of the ion occurs. The inset in figure \ref{fig:Tcurve_1c1r_P0000} shows how the internal state of the ion changes over time under irradiation with linearly polarised laser beams. All Zeeman sublevels are efficiently addressed so that no dark states arise in which the population remains trapped. Absorption events occur across the whole ion trajectory as shown in figure \ref{fig:Absevents}. Cooling transitions occur under an anti-parallel projection of the ion's velocity vector onto the laser propagation direction. Such events are indicated by red dots in figure \ref{fig:Absevents} and occur significantly more often than heating transitions with a parallel projection of the velocity onto the laser's $k$-vector (black dots). This observation highlights the importance of the Doppler shift in compensating the inhomogeneous Zeeman shifts, similarly to the dynamics in a Zeeman slower \cite{foot05a}.
\begin{figure}[!h]
    \centering
    \includegraphics[width=\linewidth]{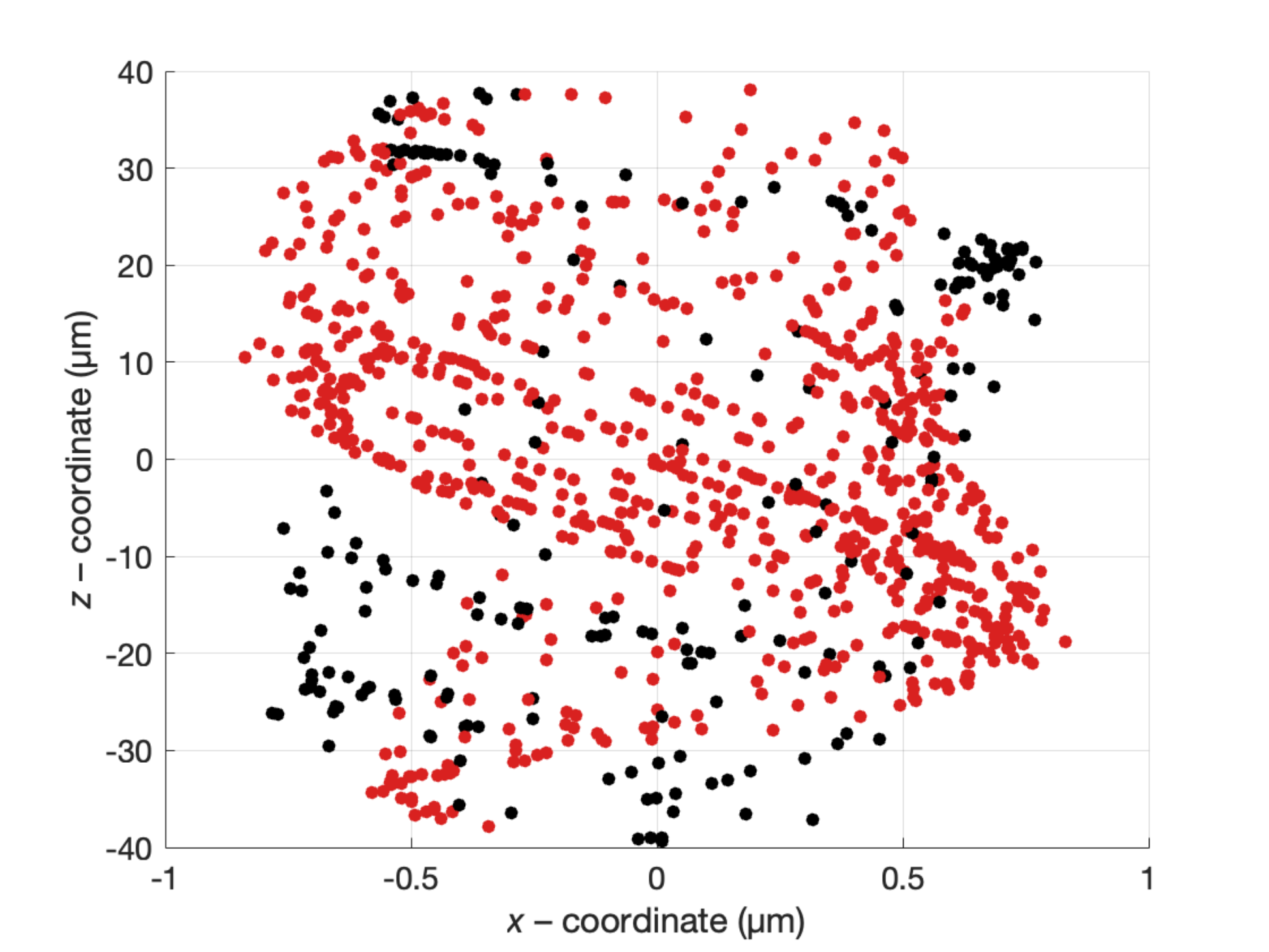}
    \caption{Locations in the lab frame at which absorption events of 397~nm photons occurred along a representative ion trajectory. Events with an anti-parallel (parallel) projection of the ion's velocity vector onto the laser propagation direction are depicted by red (black) dots. The initial velocity vector of the ion was [0.68, -0.89, -24.92]~m/s.}
    \label{fig:Absevents}
\end{figure}\\
When using linear laser polarisations (red and yellow-dashed traces in figure \ref{fig:Tcurve_1c1r_P0000} (b)), laser cooling in the hybrid trap is only slightly less efficient than in the absence of a $B$-field (green traces in figure \ref{fig:Tcurve_1c1r_P0000} (b)). Changing the orientation of the linear polarisation vector has no noticeable effect on the cooling efficiency. However, using circularly polarised laser beams results in less efficient cooling, as depicted by the blue traces in figure \ref{fig:Tcurve_1c1r_P0000}. We attribute this to the fact that the magnetic field in a quadrupolar magnetic trap points predominantly parallel or anti-parallel to the laser beams along the axial direction (see figure \ref{fig:Bfield} in the appendix). For circular polarisation, absorption events occurring on that axis are thus approximately dominated by either pure $\sigma^+$ or $\sigma^-$ transitions depending on the position of the ion with respect to the trap centre. In this scenario, specific Zeeman components of the (3d)~$^2$D$_{3/2}$ level become "dark" states that cannot be addressed by the repumper laser beams. Thus, population remains trapped impairing cooling.
Linearly polarised beams, on the other hand, can drive either $\sigma^+$ or $\sigma^-$ transitions, regardless of the position of the ion on the axis. The effect of circular polarisation diminishes for an ion with larger motional components in the radial plane (light blue curve in figure \ref{fig:Tcurve_1c1r_P0000}) for which the $B$-field deviates more strongly from a parallel alignment with the laser beam at increasing distance from the axis. Such a configuration produces more diverse projections of the laser polarisation onto the atomic frame also in the case of circular polarisation.
\begin{figure}[!h]
    \centering
    \includegraphics[width=\linewidth]{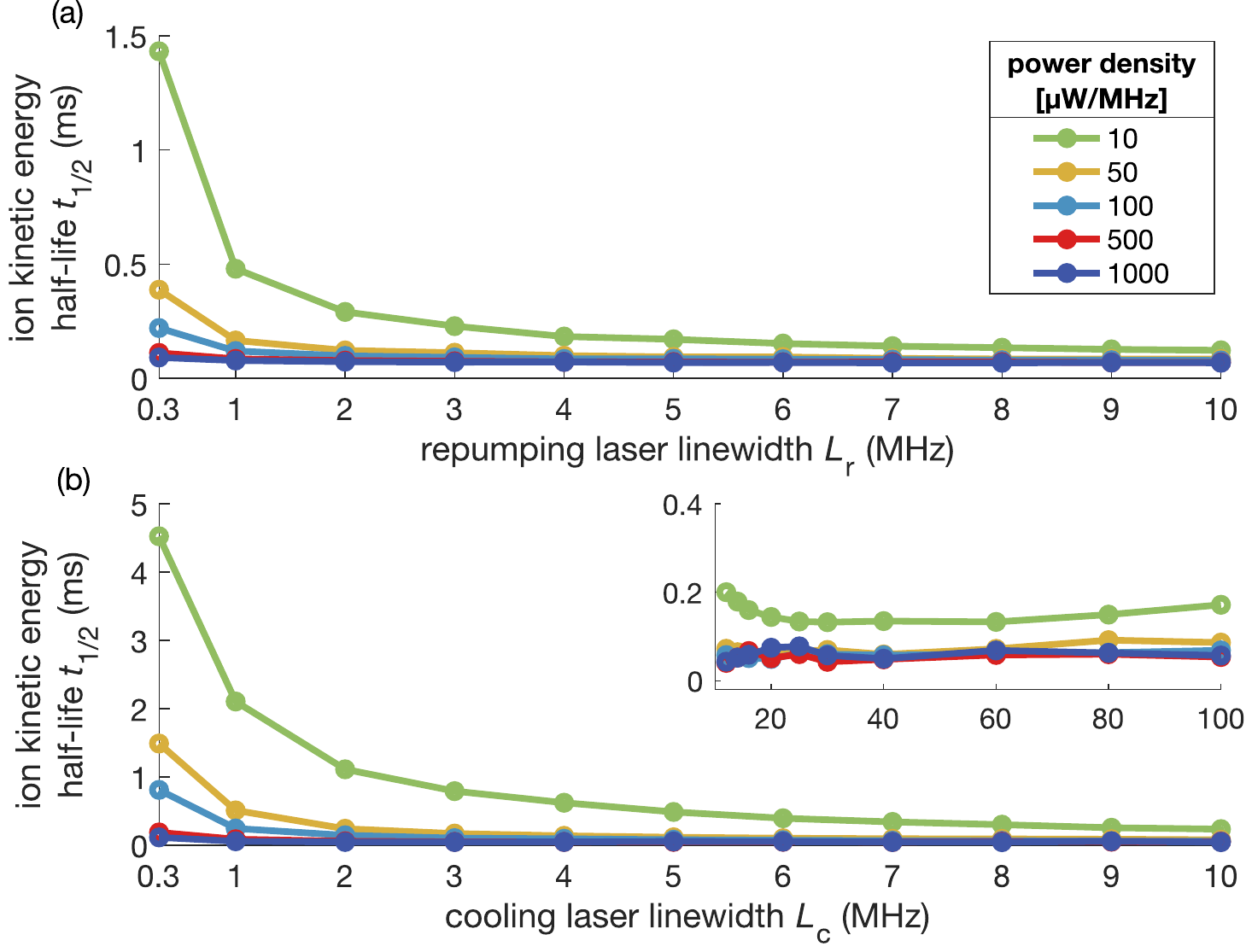}
    \caption{Dependency of the energy half-life on the linewidth of the cooling laser $L_\text{c}$ and of the repumper laser $L_\text{r}$ for different laser power spectral densities. The linewidth of the (a) cooling and (b) repumper laser remained fixed while the (a) repumper and (b) cooling laser linewidth was varied. For each specific linewidth, the saturation parameter was chosen to produce the quoted laser power spectral densities. A decrease of the half-life with increasing linewidth is only observable for low power densities indicating that power broadening is sufficient to cover the required spectral width for efficient laser cooling. Detunings of $\delta_\text{c}=-30$ MHz and $\delta_\text{r}=-10$ MHz were assumed for the fixed-frequency lasers.}
    \label{fig:SL}
\end{figure}\\
In figure \ref{fig:SL}, the ion kinetic energy half-life is plotted for different laser power spectral densities as a function of the cooling laser linewidth $L_\text{c}$ while the repumper laser linewidth $L_\text{r}$ is fixed (panel (a))  and vice versa (panel (b)). Increasing the power density at a fixed linewidth results in shorter half-lives. When keeping the power density fixed, an increase in linewidth reduces the half-life only for linewidths up to a few MHz. This effect is hardly visible for the higher power densities shown here which we attribute to power broadening.

\subsection{\label{sec:Offset}Effect of trap misalignments}
So far, an ideal experimental configuration has been studied in which the centre of the ion trap and the centre of the magnetic trap are perfectly overlapped. Experimentally, this condition is often challenging to achieve and a finite displacement between the two trap centres is difficult to avoid. If an offset between the trap centres exists, an ion oscillating around the ion-trap centre does not experience a symmetric quadrupolar field which vanishes at the centre of its trajectory. Instead, depending on the magnitude of the offset, the $B$-field across the ion's trajectory resembles more of a linear gradient.
\begin{figure}[!h]
    \centering
    \includegraphics[width=\linewidth]{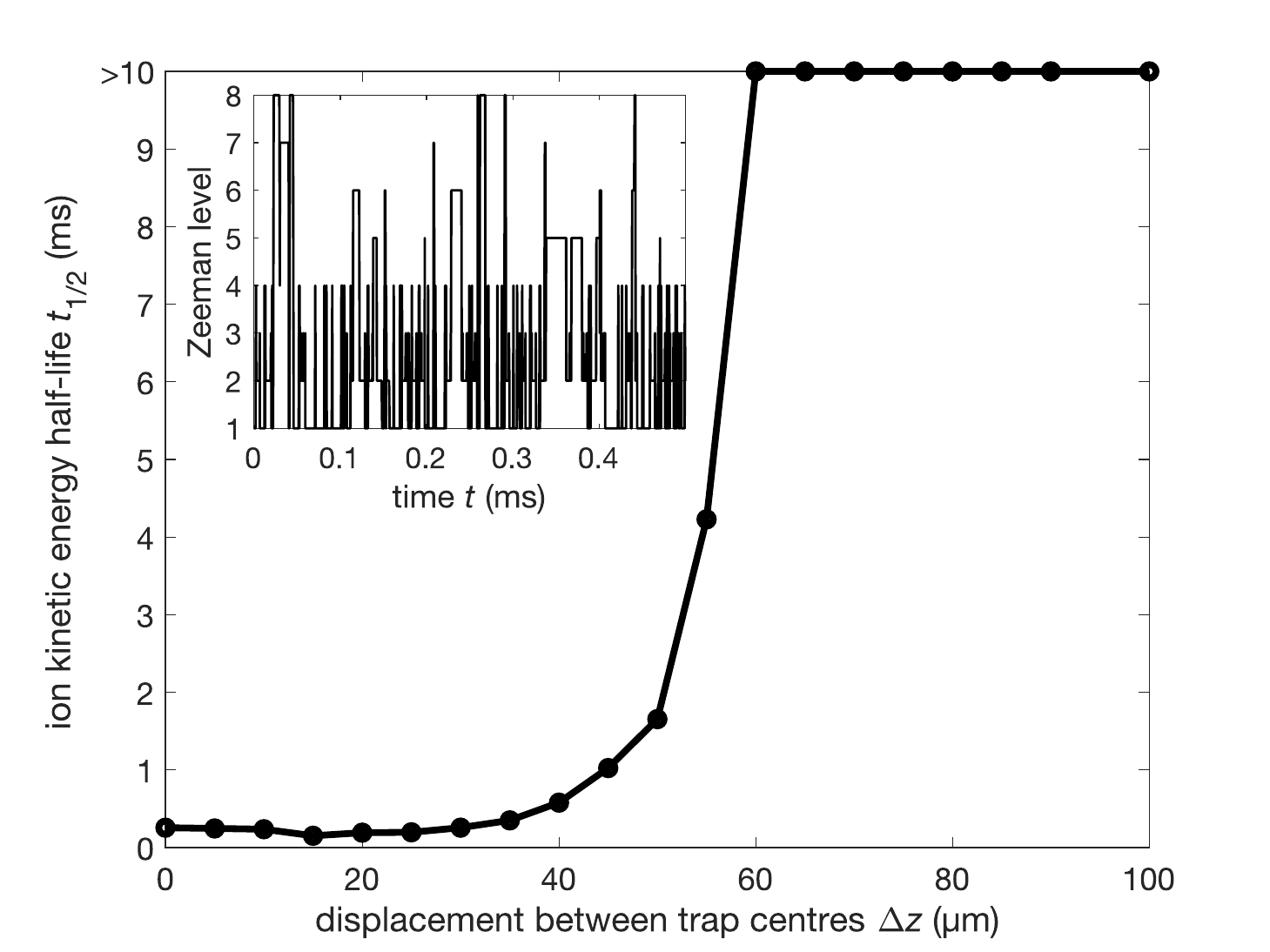}
    \caption{Effect of a misalignment between the RF and magnetic traps. An increase in distance between the two trap centres along the positive axial direction results in an increase in the energy half-life for laser parameters that were found for optimal cooling in the case of a perfect overlap of the traps. The inset shows the population of the eight Zeeman sublevels over the course of an ion trajectory at an offset of 80~$\umu$m (see figure \ref{fig:Calcium} for labelling of the levels). Compared to the ideal case shown in figure \ref{fig:Tcurve_1c1r_P0000} (a), much fewer absorption events occur. The initial velocity vector of the ion was [0.68, -0.89, -24.92] m/s.}    
    \label{fig:offsetZ}
\end{figure}\\
The effect of such a trap displacement on the cooling efficiency is shown in figure \ref{fig:offsetZ}. An increase in the offset between the two trap centres results in an increase of the half-life for one cooling and one repumper laser at parameters that were found optimal for the ideal hybrid trap (see figure \ref{fig:Tcurve_1c1r_P0000} (a)). A marked deterioration of the half-life was found to occur already at displacements exceeding 40~$\umu$m. The number of absorption events over the same period of time also decreases with the magnitude of the trap displacement, which suggests that the laser detunings need to be adjusted. However, scanning the detuning and the polarisation of the cooling laser beam at a fixed finite offset was not found to yield conditions for efficient cooling. We attribute this to the range of Zeeman splittings sampled in this scenario which are too large in order to be efficiently addressed by a single monochromatic cooling laser beam. To mitigate this effect, a second cooling laser beam can be added and the detunings of the two cooling lasers can be adjusted individually. 
\begin{figure}[!h]
    \centering
\includegraphics[width=\linewidth]{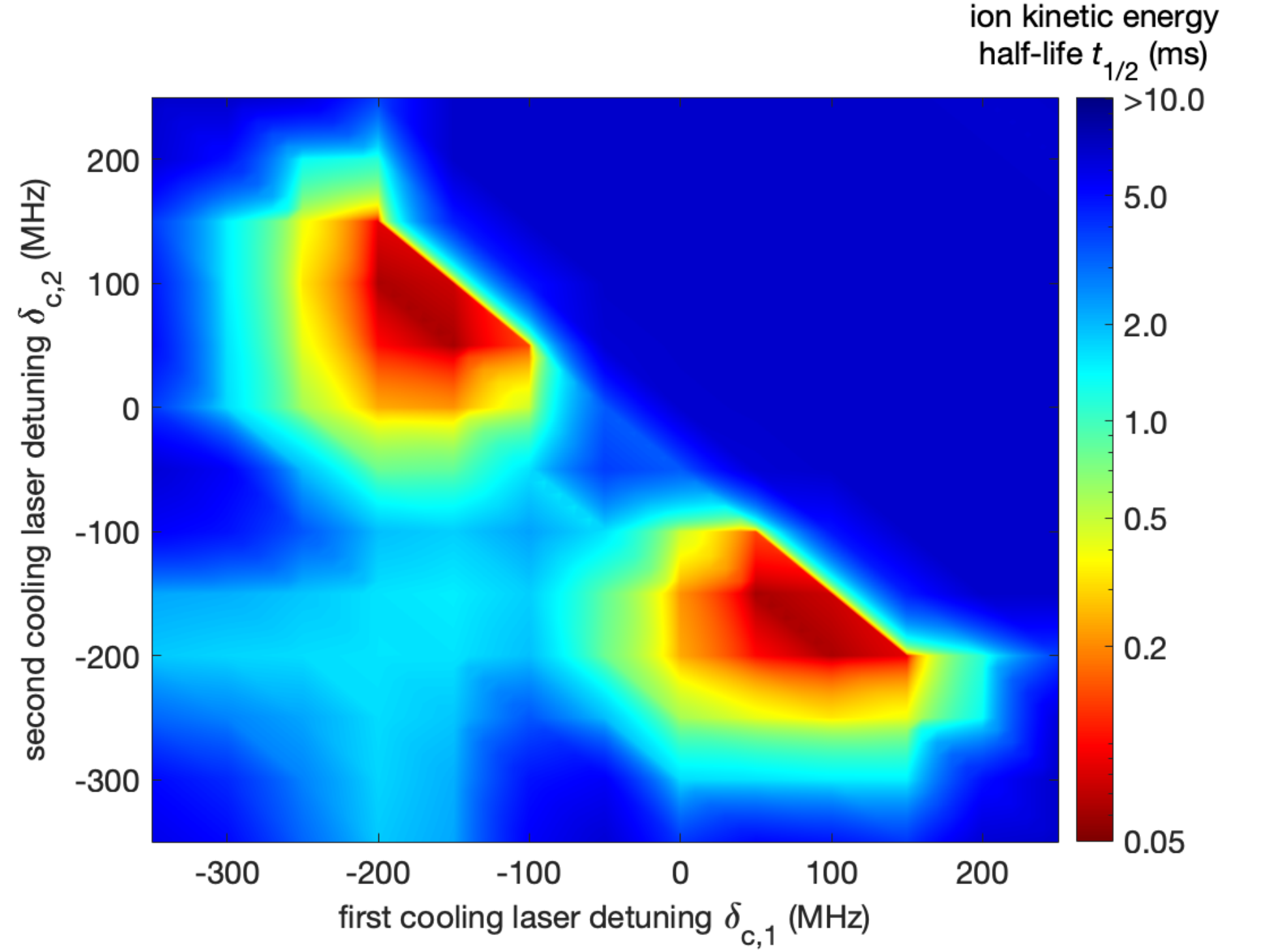}
    \caption{Ion kinetic energy half-lives $t_{1/2}$ (in ms) for different detunings of two co-propagating cooling lasers at a relative trap displacement of 80~$\umu$m along the axial direction. Efficient laser cooling at such a trap displacement can be recovered by introducing a second cooling laser and setting the detunings of the two lasers such that they deplete the two $^2$S$_{1/2}$ Zeeman sublevels (e.g. $\delta_\text{c,1}=-150$ MHz and $\delta_\text{c,2}=50$ MHz). Only the cooling transition $^2$S$_{1/2} - ^2$P$_{1/2}$ was considered in this figure.}
    \label{fig:Dscan_offset}
\end{figure}\\
\begin{figure}[!h]
    \centering
    \includegraphics[width=\linewidth]{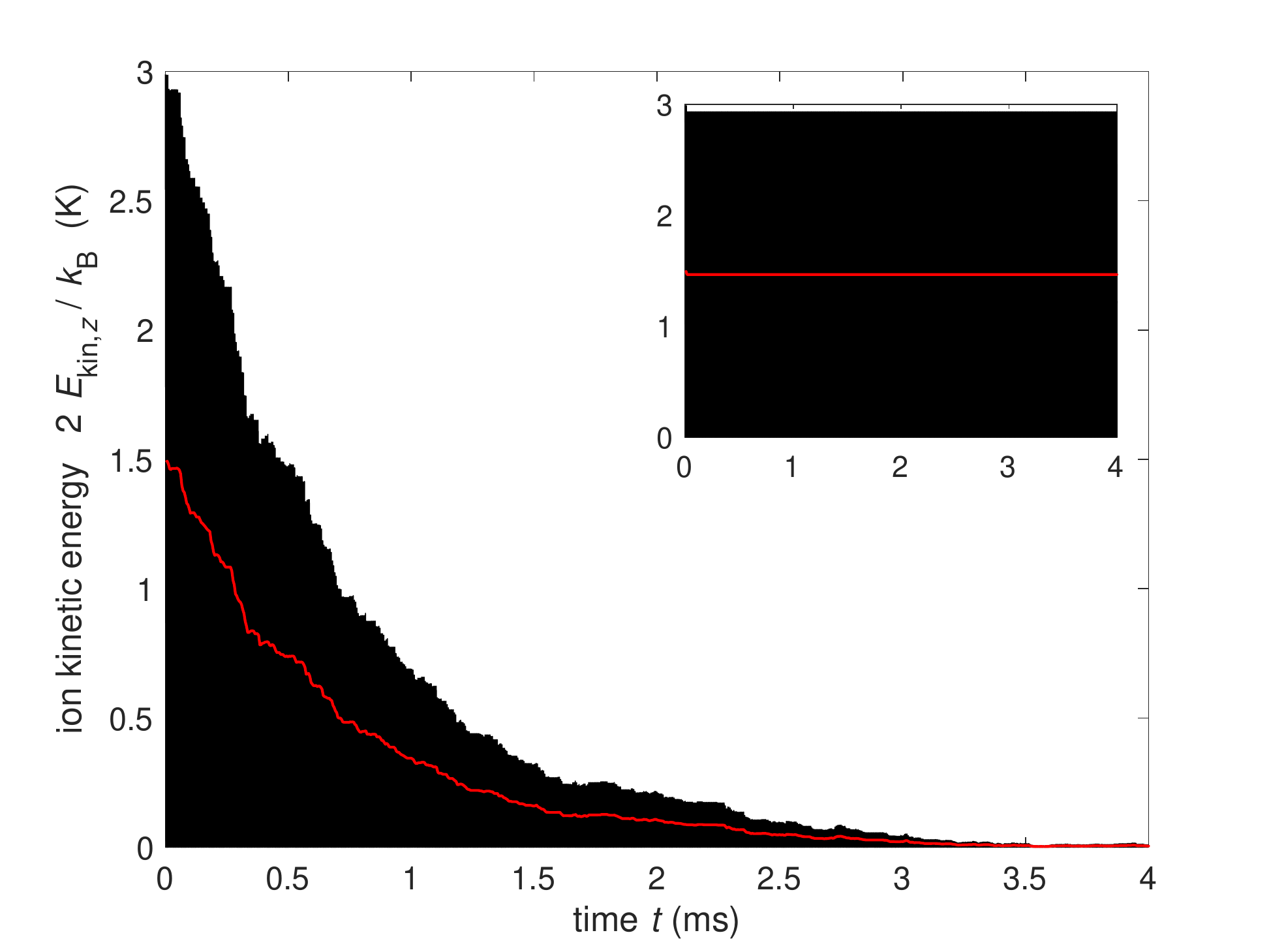}
    \caption{Kinetic energy along the axial direction of the ion motion (black trace) at a trap offset of 80~$\umu$m for two co-propagating cooling lasers and a single repumper laser. The red trace depicts the average over the secular motion. The detunings of the two cooling lasers were $\delta_\text{c,1}=-150$ MHz and $\delta_\text{c,2}=50$ MHz. The repumper laser exhibited a detuning of $\delta_\text{r}=0$ MHz and linear polarisation aligned with the $x$ axis. No cooling occurs if the repumper laser has circular polarisation which is shown in the inset for left circular polarisation.}
    \label{fig:Repumper_offset}
\end{figure}
To illustrate this point, we reduced the calcium ion to a hypothetical 4-level system consisting of the $^2$S$_{1/2}$ and the $^2$P$_{1/2}$ Zeeman sublevels and scanned the detunings of the two cooling lasers. The results of these simulations are shown in figure \ref{fig:Dscan_offset}. It can be seen that efficient cooling can be retrieved by adding a second cooling laser and choosing the detunings such that transitions out of both magnetic components of the $^2$S$_{1/2}$ level are adequately addressed. \\
Considering again the full eight-level ion and choosing the detunings $\delta_{c,1} = -150$ MHz and $\delta_{c,2} = 50$ MHz for the two cooling lasers, the effect of the repumper laser is studied. Figure \ref{fig:Repumper_offset} shows the kinetic energy along the axial direction for a single repumper laser beam with linear polarisation parallel to the $x$ axis. Despite large Zeeman splittings, a single repumper laser with linear polarisation is sufficient to address the $^2$D$_{3/2} - ^2$P$_{1/2}$ transitions without producing dark states. According to figure \ref{fig:Dscan_1c1r_P0000}, the population in the $^2$D$_{3/2}$ levels can be sufficiently repumped even when the laser is detuned by several tens of MHz. In contrast to the perfectly overlapped hybrid trap, where circular laser polarisation does not necessarily always lead to the occurrence of dark states as a consequence of the inhomogeneity of the magnetic field, the field is much less inhomogeneous in the presence of a trap offset. The $B$-field vector components along the axial direction sampled by the ion all have the same sign. While linear polarisation still drives both $\sigma_+$ and $\sigma_-$ transitions, circular polarisation can only drive one of the two $\sigma$-transitions across the whole ion trajectory in this situation. This results in dark states $^2$D$_{3/2}$ Zeeman sublevels which prohibit laser cooling as shown in the inset.\\
Besides large Zeeman splittings across the whole ion trajectory, misalignments of the two trap centres also have pronounced effects on the characteristics of the ion motion. Figure \ref{fig:IonMotion} shows the trajectory of an ion in (a) \& (c) the ideal case with zero displacement and (b) \& (d) with a displacement of 80 $\umu$m between the centre of the ion and the magnetic trap along the axial direction. The three axes in a perfectly overlapped hybrid trap are only very weakly coupled for a single ion such that its trajectory in the radial plain remains in the direction of its initial velocity over the first hundreds of microseconds. When the two trap centres are displaced, the $B$-field not only becomes less inhomogeneous but also stronger in magnitude. In this situation, the magnetic Lorentz force leads to motional components in the radial plain that resemble cyclotron motion in Penning traps. It can thus be surmised that a hybrid trap with a very large offset between the trap centres will exhibit features reminiscent of a Penning trap.
\begin{figure}[!h]
    \centering
    \includegraphics[width=\linewidth]{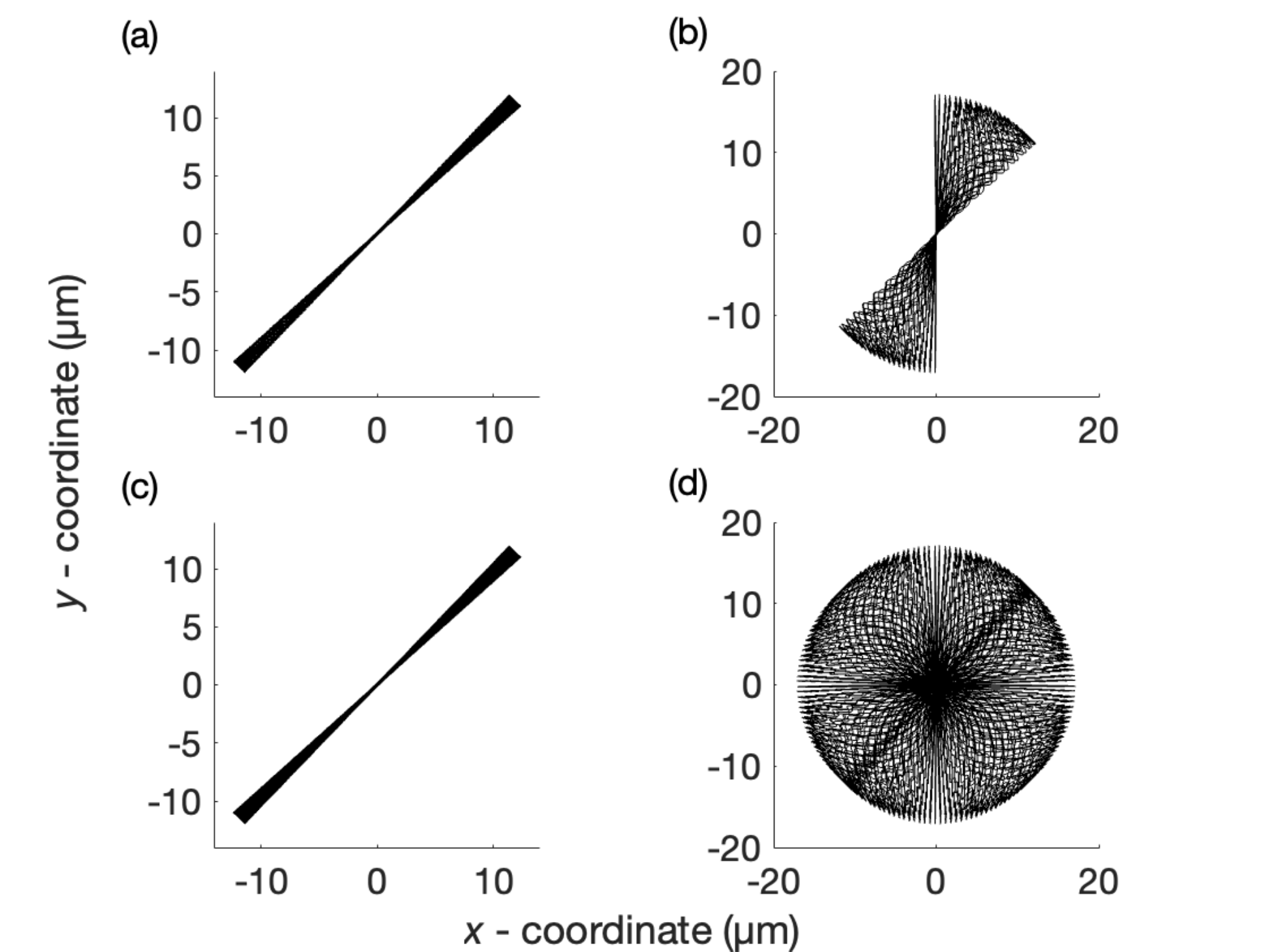}
    \caption{Trajectory of an ion in the hybrid trap (a) \& (b) over the course of 100~$\umu$s and (c) \& (d) over 400~$\umu$s for (a) \& (c) a perfect overlap between the ion and magnetic traps and for (b) \& (d) a displacement of 80~$\umu$m along the positive axial direction between the trap centres of the ion and magnetic trap. While the oscillatory motion of the ion in the radial plain remains in the direction of the initial velocity vector over the course of hundreds of microseconds in a perfectly overlapped hybrid trap, the trajectory in the presence of a trap centre displacement shows signatures of cyclotron motion reminiscent of Penning traps. The initial velocity vector of the ion was [14.63, 15.18, 13.33]~m/s.}
    \label{fig:IonMotion}
\end{figure}

\newpage
\section{\label{sec:Conclusion}Conclusion}
Using an eight-level rate equation treatment for the state populations combined with a molecular-dynamics framework, we studied theoretically the population and motional dynamics of a single Ca$^+$ ion in the combined quadrupolar magnetic and electric radiofrequency fields of a hybrid trap under the action of cooling and repumper laser beams. Laser cooling of a single cold ion was found to work efficiently in the hybrid trap with laser parameters commonly used for the Doppler cooling of trapped ions in the absence of a $B$-field. If slight displacements between the centres of the ion trap and of the quadrupolar magnetic field are introduced, as can be expected to be the case in real experimental implementations, the ion motion starts to increasingly sample linear $B$-field gradients rather than a quadrupolar $B$-field configuration. In this scenario, the ion motion assumes cyclotron-like components, and the laser cooling efficiency was found to markedly deteriorate but can be recovered by employing additional cooling laser beams.

\section*{Acknowledgments}
We acknowledge financial support from the Swiss National Science Foundation, grant nrs. 200021\_204123 and TMAG-2\_209193 as well as the University of Basel.



\section*{Data availability}
The codes used to produce the findings of this study are available on the Zenodo repository with the identifier DOI: 10.5281/zenodo.7445811.\\


\bibliographystyle{tfo}
\bibliography{GroupBib,NewReferences}

\appendix
\newpage
\section{\label{sec:App_REvsOBE}Rate equations vs. optical Bloch equations}

\begin{figure}[!h]
    \centering
    \includegraphics[width=\linewidth]{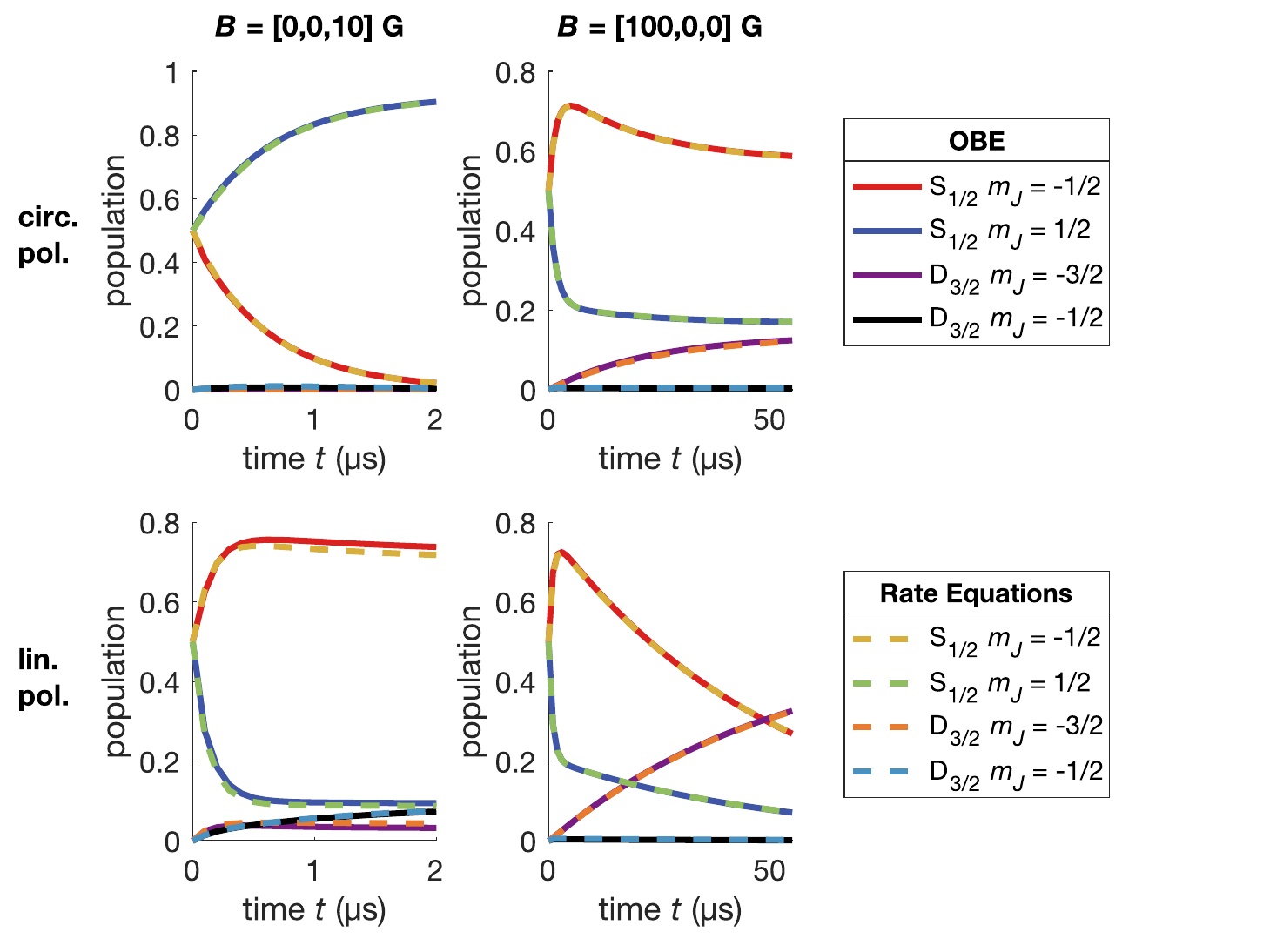}
d    \caption{The time evolution of selected state populations of a calcium ion at rest. Solutions obtained via rate equations (dashed lines) match with the solutions of optical Bloch equations (solid lines) for circular (first row) and linear (second row) laser polarisation at different $B$-field strengths and directions. One cooling and one repumper laser were assumed to propagate along the $z$ axis with saturation parameters of $S_{\text{c}}=1.96$, $S_{\text{r}}=105.75$, detunings of $\delta_{\text{c}}=-30$ MHz, $\delta_{\text{r}}=-10$ MHz and linewidths of $L_{\text{c}}=0.3$ MHz, $L_{\text{r}}=0.3$ MHz.}
    \label{fig:REvsOBE_1c1r}
\end{figure}

\begin{figure}[!h]
    \centering
    \includegraphics[width=\linewidth]{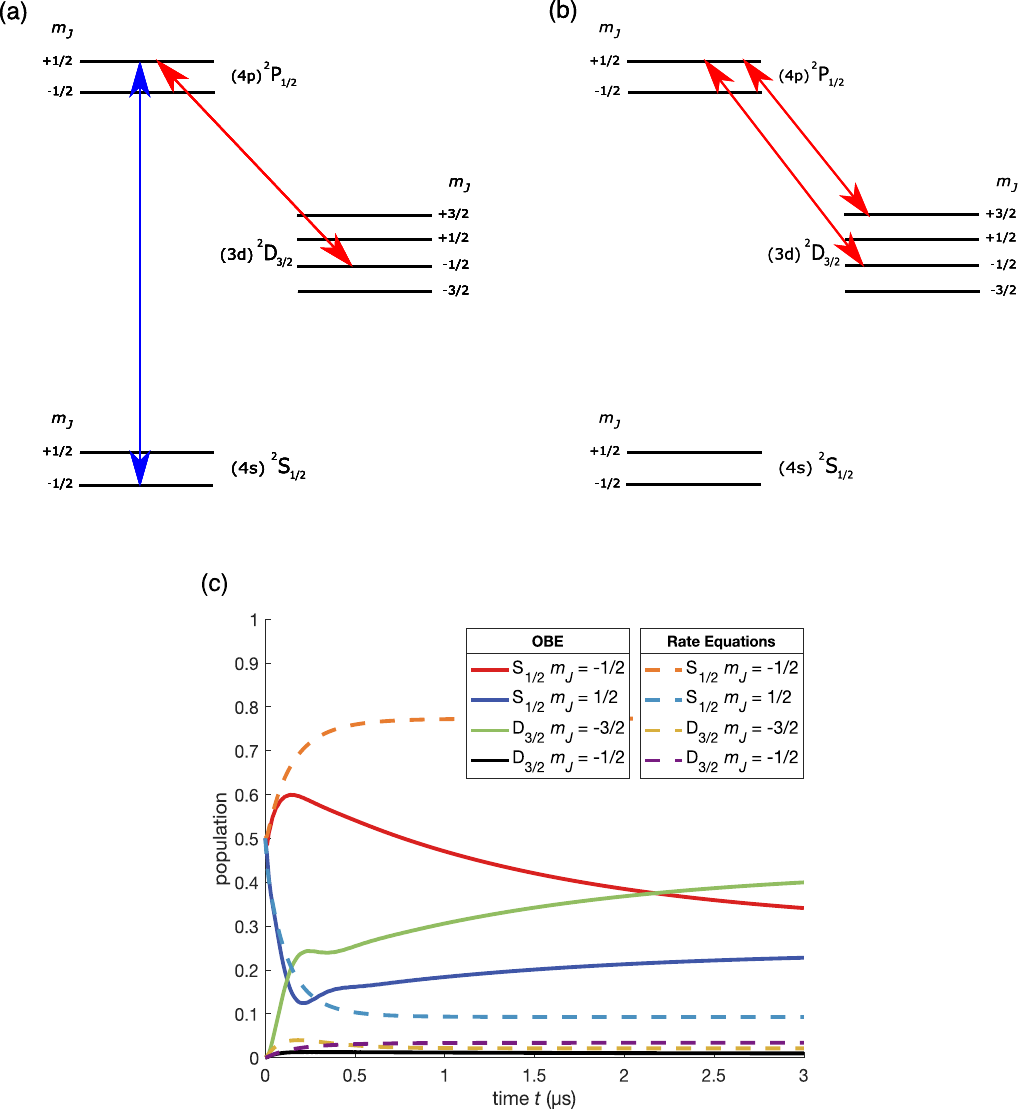}
    \caption{Two scenarios for CPT conditions in an eight-level system in (a) S-P-D and (b) D-P-D lambda systems.  (c) When the CPT conditions are met, rate equations and OBE give different solutions as shown with the example of two counter-propagating cooling and two counter-propagating repumper laser beams corresponding to the scenario shown in (a). The lasers propagate in the same direction as a 10 G $B$-field and have saturation parameters of $S_{\text{c}}=1$, $S_{\text{r}}=100$, detunings of $\delta_{\text{c}}=-30$ MHz, $\delta_{\text{r}}=0$ MHz, linewidths of $L_{\text{c}}=0$ MHz, $L_{\text{r}}=0$ MHz, and circular polarisation.}
    \label{fig:CPT}
\end{figure}

In order to verify the rate equation (RE) model and identify its limitations, we compared the computed population dynamics to the solutions of optical Bloch equations (OBE)  \cite{oberst99a,gingell10a,dijck20a}. In this treatment, we solved the Liouville equation 
\begin{equation}
\label{eq:Liouville}
\frac{\text{d}\rho}{\text{d}t} = -\frac{\text{i}}{\hbar} [H,\rho] + L(\rho)
\end{equation} 
for the eight-level system in the calcium ion and computed the evolution of the density matrix $\rho$ for different laser parameters and $B$-fields. The Hamiltonian $H=H_\text{A}+H_\text{I}$ consists of the unperturbed atomic Hamiltonian $H_\text{A}$ and the interaction Hamiltonian $H_\text{I}$. Dissipative processes like spontaneous emission and finite laser linewidths are included in the Lindblad operator $L(\rho)$. In figure \ref{fig:REvsOBE_1c1r}, the diagonal elements of $\rho$ according to equation (\ref{eq:Liouville}) are compared to the solutions of equations (1) and (2) in the main text. Two Zeeman sublevels of each the (4s)~$^2$S$_{1/2}$ and (3d)~$^2$D$_{3/2}$ states are plotted and different laser polarisations and $B$-field orientations are explored. Rate equations and OBE were found to match almost perfectly. In particular, the population follows the selection rules for different polarisations with respect to the $B$-field direction, and slower evolution is observed for stronger $B$-field strengths due to larger Zeeman splittings.\\
However, there are situations in which larger discrepancies between the rate-equation and OBE treatments can be observed. For instance, coherent population trapping (CPT) occurs if two lasers that drive a lambda system exhibit the same detuning \cite{gray78a}. In this situation, a superposition state composed of the two ground states of the lambda system is generated. This effect arises due to the coherences between atomic states which are modelled by the OBE but not by rate equations. In the 8-level calcium ion, several lambda systems can be found. CPT effects can either occur between (4s)~$^2$S$_{1/2}$ and (3d)~$^2$D$_{3/2}$ Zeeman sublevels or between two (3d)~$^2$D$_{3/2}$ Zeeman sublevels if two repumper lasers are used. Two examples of lambda systems that can sustain CPT in the calcium ion are shown in figure \ref{fig:CPT}. The population evolution calculated by rate equations and by OBE is also shown for laser parameters at which CPT occurs.\\
While the detuning depends only on the laser frequency and the Zeeman shift for an ion at rest, the detuning of a moving ion depends also on the Doppler shift. In trajectory simulations of an ion in an inhomogeneous $B$-field, the ion constantly changes position and velocity and thus experiences different Zeeman and Doppler shifts at each time step. Thus, despite being present in our system, we regard CPT effects as negligible due to the motional dynamics. Overall, we conclude that the rate equation model presented in section 2.2 of the main text is appropriate to study the population dynamics in the present case.\\

\newpage
\section{\label{sec:App_ExpSetup} Assumed experimental setup}
\begin{figure}[!h]
    \centering
    \includegraphics[width=\linewidth]{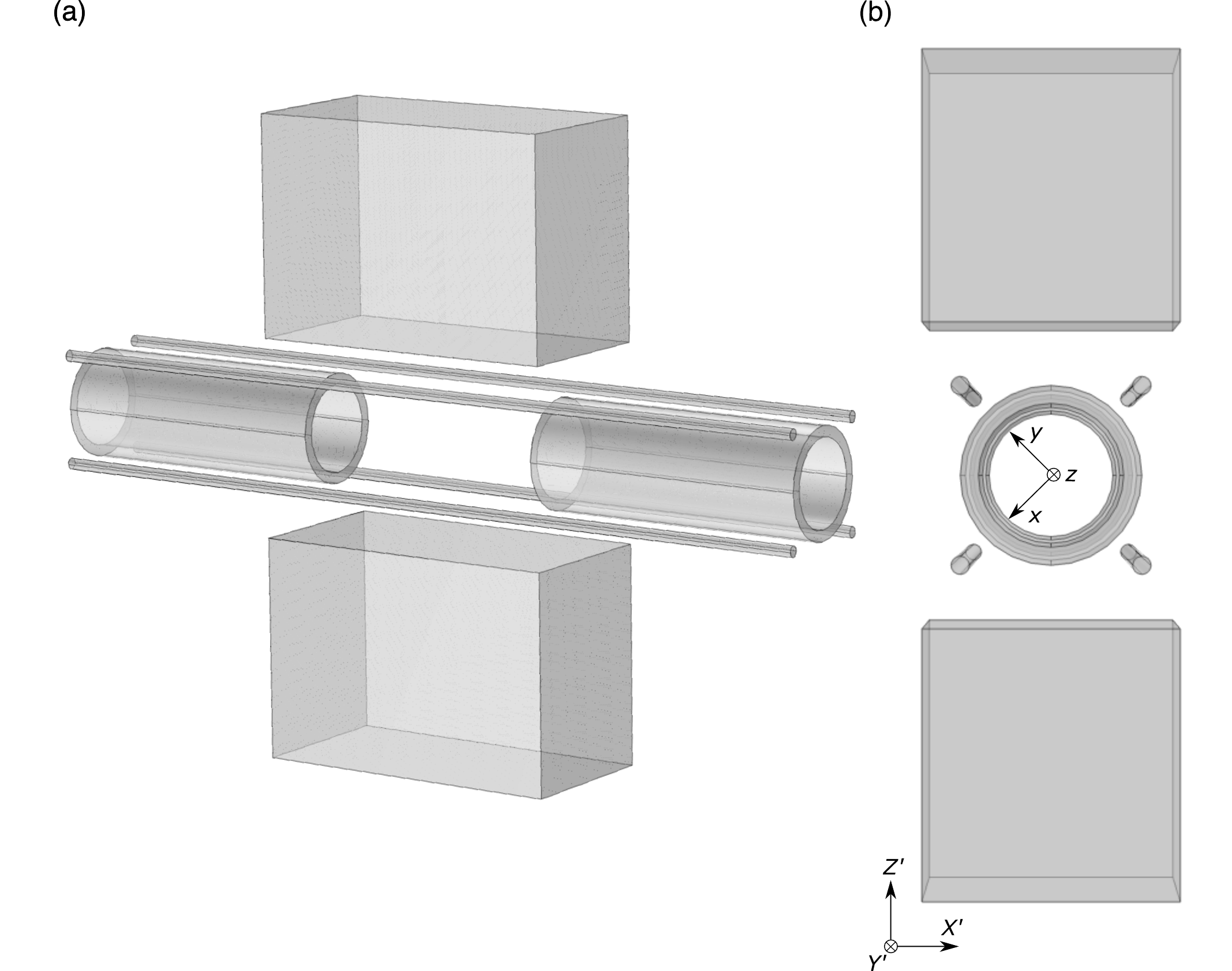}
    \caption{Schematic of the assumed prototypical hybrid trap setup from (a) a side view and (b) a frontal view. The hybrid trap is composed of a linear Paul trap and two bar magnets. The laboratory frame for the ion ($x,y,z$) and a coordinate system used to describe the magnetic fields  ($X', Y', Z'$) are shown in (b). See the text for the exact definitions of the coordinate systems.}
    \label{fig:ExpSetup}
\end{figure}

\begin{figure}[!h]
    \centering
    \includegraphics[width=\linewidth]{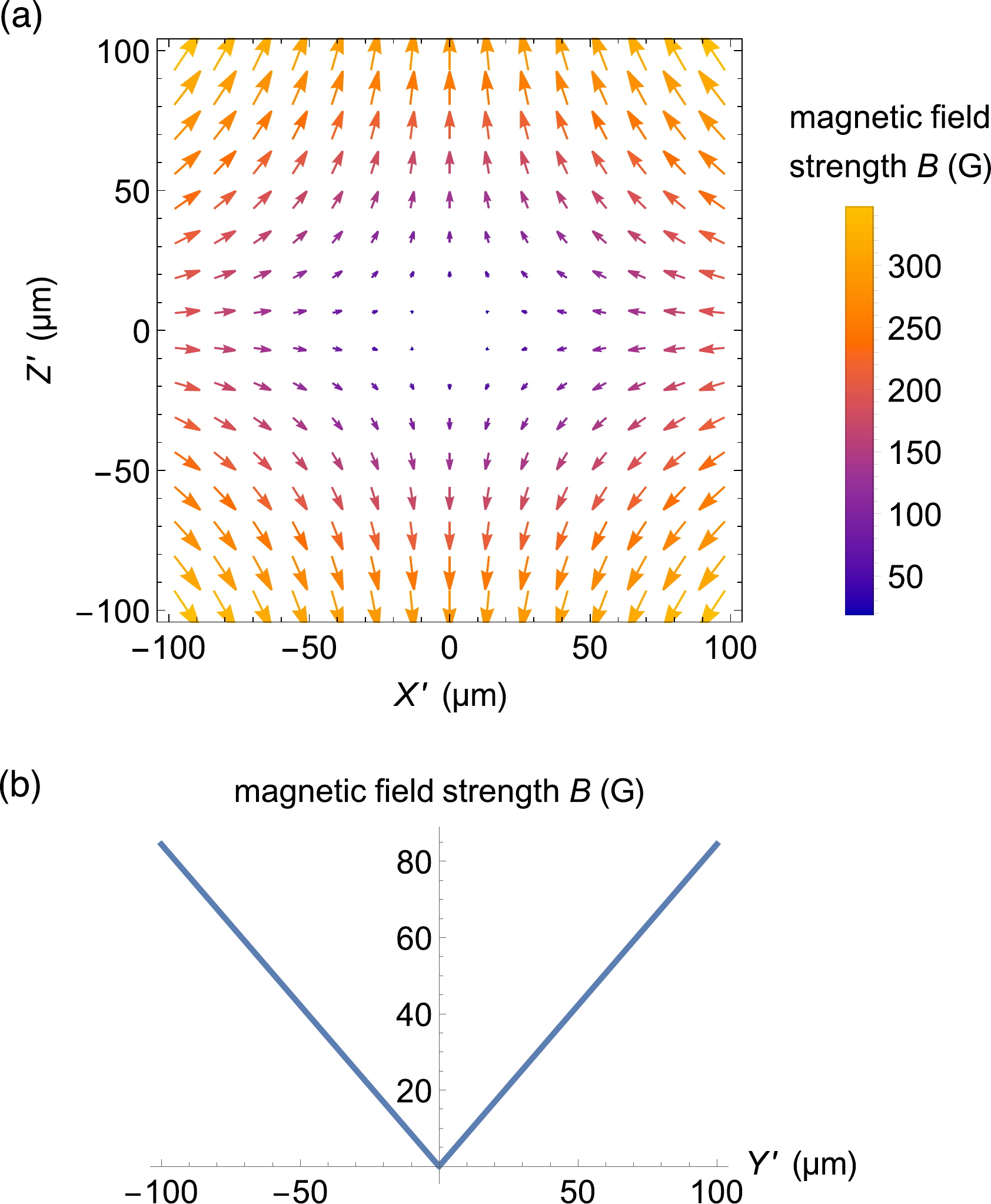}
    \caption{Quadrupolar $B$-field produced by the magnetic trap. (a) Vector field in the radial plain at $Y'=0$ and (b) $B$-field strength along the axial direction of the ion trap.}
    \label{fig:Bfield}
\end{figure}

Figure \ref{fig:ExpSetup} shows a schematic of a prototypical hybrid trap assumed in the present simulations. The linear RF trap is composed of four cylindrical electrodes for the application of radio-frequency fields and two endcap electrodes for static fields to close the trap along the axial direction. The magnetic trap overlapped with the ion trap is assumed to be composed of two bar magnets. The $x$ and $y$ axes point toward two RF electrodes while the $z$ axis lies parallel to the trap axis as indicated. This coordinate system defines the lab frame used in the equations of motion of the ion (equation (5) in the main text) and has its origin at the centre of the hybrid trap. Special care must be taken to match the lab frame with the frame given by the $k$-vector of a laser beam and with the atomic frame (see section 2.2 in the main text). Moreover, we used a coordinate system $X'Y'Z'$ in which the axes coincidence with the sides of the bar magnets ($X' = -(x+y)/\sqrt{2},\ Y' = z,\ Z' = (y-x)/\sqrt{2}$) for the magnetic trap. Figure \ref{fig:Bfield} (a) shows the magnetic vector field in the radial plain at $Y'=z=0$ and (b) the $B$-field strength in the axial direction in this coordinate system according to the analytical expressions given in reference \cite{engel-herbert05a}. The $B$-field exhibits gradients $\delta B / \delta X' = 205$~T/m, $\delta B / \delta Y' = 85$~T/m, and $\delta B / \delta Z' = 290$~T/m with respect to the trap centre.\\
\\
\\

\newpage
\section{\label{sec:App_MorePlots}Simulation results at different initial conditions}

\begin{figure}[!h]
\centering
\includegraphics[width=\linewidth]{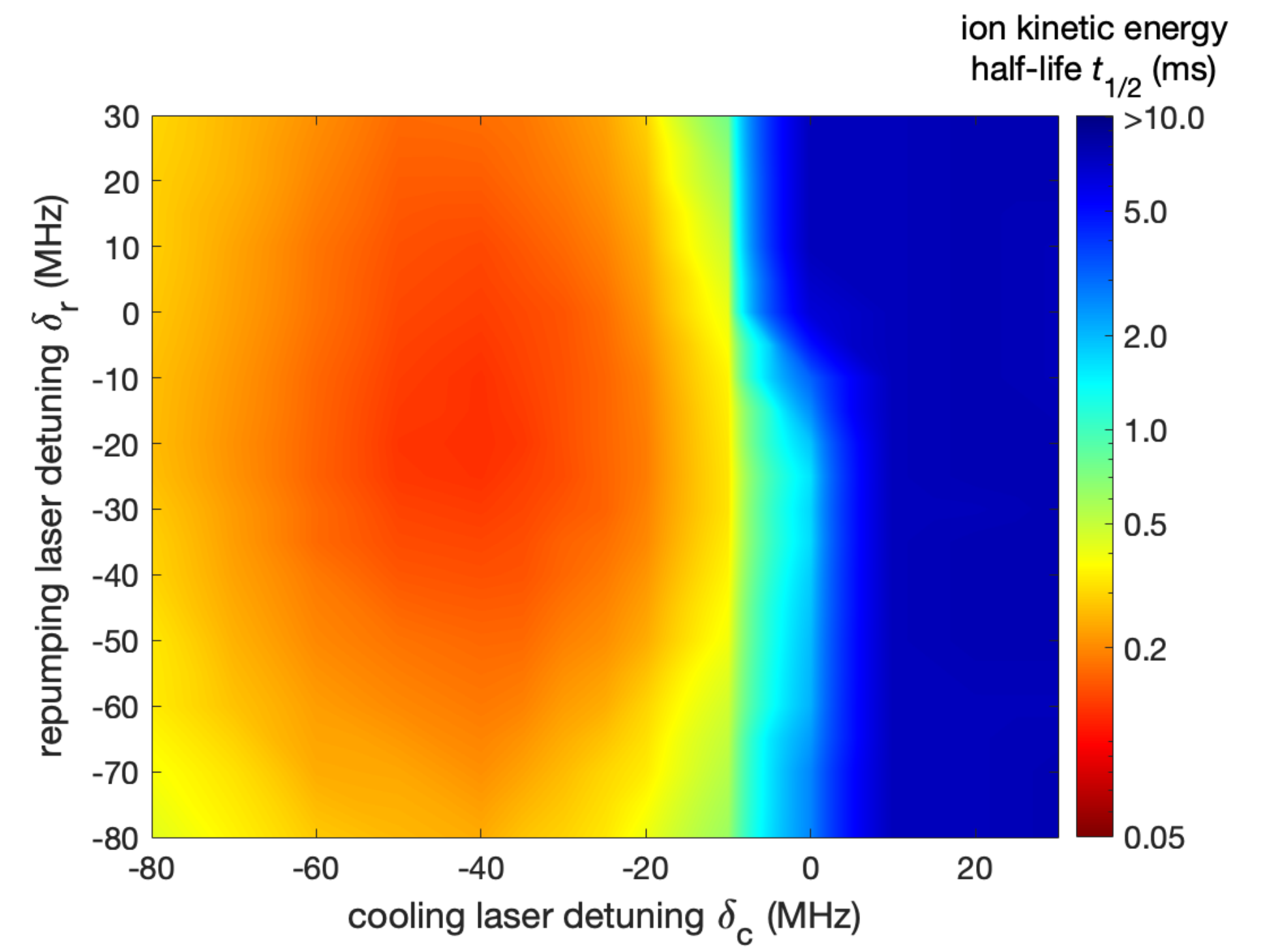}
\caption{Ion kinetic energy half-lives $t_{1/2}$ at different laser detunings for one cooling and one repumper laser beam propagating along the axial direction of the hybrid trap. The initial position of the ion is 10~$\umu$m from the trap centre. Similar results are obtained as for an ion with an initial position at the trap centre (compare with figure 5 of the main text). The laser parameters used here were $S_\text{c}=1.96$, $S_\text{r}=105.75$, $L_\text{c}=0.3$ MHz, $L_\text{r}=0.3$ MHz and linear polarisation along the $x$ axis.}
\label{fig:Dscan_position10um}
\end{figure}

\begin{figure}[!h]
\centering
\includegraphics[width=\linewidth]{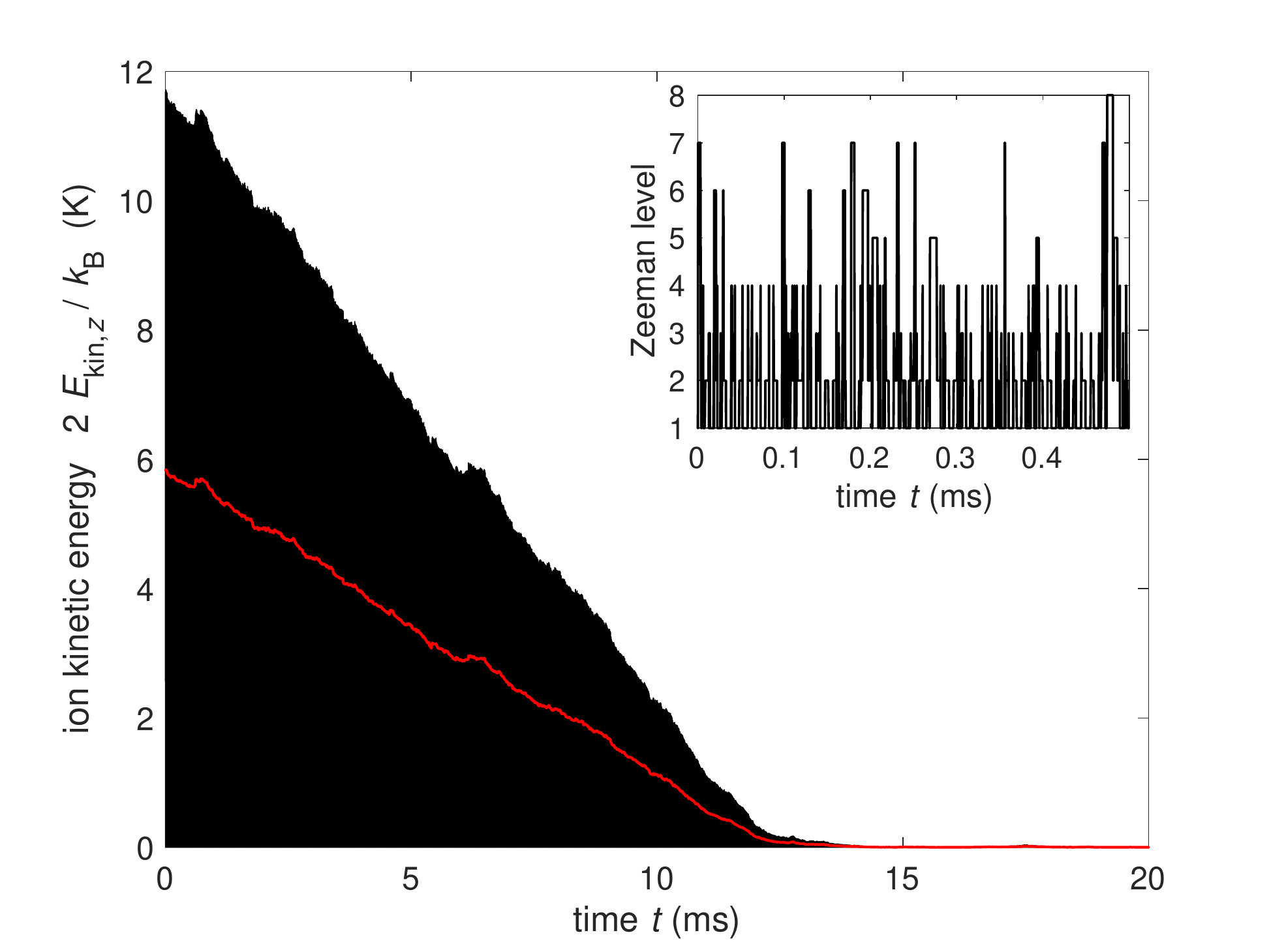}
\caption{Time evolution of the kinetic energy of an ion along the axial direction (black trace) and its average over the secular frequency (red trace). The inset shows the population of the eight Zeeman sublevels over the course of the ion trajectory for the first 0.5 ms (see figure 1 in the main text for labelling of the levels). The initial velocity vector was [-43.14, -43.96, 49.28]~m/s which corresponds to 10 K according to the equipartition theorem. The laser parameters used here were $S_\text{c}=1.96$, $S_\text{r}=105.75$, $\delta_\text{c}=-30$ MHz, $\delta_\text{r}=-10$ MHz, $L_\text{c}=0.3$ MHz, $L_\text{r}=0.3$ MHz and linear polarisation along the $x$ axis.}
\label{fig:Tcurve_10K}
\end{figure}

In the main text, the initial position of the ion was chosen at the centre of the hybrid trap. Figure \ref{fig:Dscan_position10um} shows the average of 32 simulations, each with a different initial velocity vector corresponding to a temperature of 1 K, for an initial position 10~$\umu$m from the trap centre along the positive axial direction. Qualitatively the same results as for an ion with an initial position at the trap centre are obtained (compare with figure 5 of the main text).
\\
In the main text, the initial velocity vectors of the ion were chosen such that they correspond to 1 K according to the equipartition theorem. Figure \ref{fig:Tcurve_10K} shows the kinetic energy along the axial direction for an ion in the hybrid trap with an initial velocity vector that corresponds to 10 K. Laser cooling is also possible at higher initial temperatures but, not surprisingly, more time is required.

\end{document}